\documentclass[a4paper]{report}
\usepackage[utf8]{inputenc}
\usepackage[T1]{fontenc}
\usepackage{RJournal}
\usepackage{amsmath,amssymb,array}
\usepackage{booktabs}

\providecommand{\tightlist}{%
  \setlength{\itemsep}{0pt}\setlength{\parskip}{0pt}}

\newlength{\cslhangindent}
\newlength{\csllabelwidth}
\newlength{\cslentryspacingunit} 
  {}%
  {\par}
\newenvironment{CSLReferences}[2] 
 {
  \setlength{\parindent}{0pt}
  \ifodd #1
  \let\oldpar\par
  \def\par{\hangindent=\cslhangindent\oldpar}
  \fi
  \setlength{\parskip}{#2\cslentryspacingunit}
 }%
 {}
\usepackage{calc}

\begin{document}

\sectionhead{Contributed research article}
\volume{XX}
\volnumber{YY}
\year{20ZZ}
\month{AAAA}

\begin{article}
\title{bqror: An R package for Bayesian Quantile Regression in Ordinal Models}
\author{by Prajual Maheshwari and Mohammad Arshad Rahman}

\maketitle

\abstract{%
This article describes an R package \(\CRANpkg{bqror}\) that estimates Bayesian quantile regression in ordinal models introduced in Rahman (2016). The paper classifies ordinal models into two types and offers computationally efficient, yet simple, Markov chain Monte Carlo (MCMC) algorithms for estimating ordinal quantile regression. The generic ordinal model with 3 or more outcomes (labeled \(OR_{I}\) model) is estimated by a combination of Gibbs sampling and Metropolis-Hastings algorithm. Whereas an ordinal model with exactly 3 outcomes (labeled \(OR_{II}\) model) is estimated using a Gibbs sampling algorithm only. In line with the Bayesian literature, we suggest using the marginal likelihood for comparing alternative quantile regression models and explain how to compute the same. The models and their estimation procedures are illustrated via multiple simulation studies and implemented in two applications. The article also describes several functions contained within the \(\CRANpkg{bqror}\) package, and illustrates their usage for estimation, inference, and assessing model fit.
}

\hypertarget{sec:intro}{%
\section{Introduction}\label{sec:intro}}

Quantile regression defines the conditional quantiles of a continuous dependent variable as a function of the covariates without assuming any error distribution (Koenker and Bassett 1978). The method is robust and offers several advantages over least squares regression such as desirable equivariance properties, invariance to monotone transformation of the dependent variable, and robustness against outliers (Koenker 2005; Davino, Furno, and Vistocco 2014; Furno and Vistocco 2018). However, quantile regression with discrete outcomes is more complex because quantiles of discrete data cannot be obtained through a simple inverse operation of the cumulative distribution function (\({cdf}\)). Besides, discrete outcome (binary and ordinal) modeling requires location and scale restrictions to uniquely identify the parameters (see \protect\hyperlink{sec:QROrdinal}{Section 2} for details). Kordas (2006) estimated quantile regression with binary outcomes using simulated annealing, while Benoit and Van den Poel (2010) proposed Bayesian binary quantile regression where a working likelihood for the latent variable was constructed by assuming the error follows an asymmetric Laplace (AL) distribution (Yu and Moyeed 2001). The estimation procedure for the latter is available in the \CRANpkg{bayesQR} package of R software (Benoit and Van den Poel 2017). A couple of recent works on Bayesian quantile regression with binary longitudinal (panel) outcomes are Rahman and Vossmeyer (2019) and Bresson, Lacroix, and Rahman (2021). Extending the quantile framework to ordinal outcomes is more intricate due to the difficulty in satisfying the ordering of cut-points while sampling. Rahman (2016) introduced Bayesian quantile analysis of ordinal data and proposed two efficient MCMC algorithms. Since Rahman (2016), ordinal quantile regression has attracted some attention which includes Alhamzawi (2016), Alhamzawi and Ali (2018), Ghasemzadeh, Ganjali, and Baghfalaki (2018), Rahman and Karnawat (2019), Ghasemzadeh, Ganjali, and Baghfalaki (2020), and Tian et al. (2021).

Ordinal outcomes frequently occur in a wide class of applications in economics, finance, marketing, and the social sciences. Here, ordinal regression (e.g.~ordinal probit, ordinal logit) is an important tool for modeling, analysis, and inference. Given the prevalence of ordinal models in applications and the recent theoretical developments surrounding ordinal quantile regression, an estimation package is essential so that applied econometricians and statisticians can benefit from a more comprehensive data analysis. At present, no statistical software (such as R, MATLAB, Python, Stata, SPSS, and SAS) have any package for estimating quantile regression with ordinal outcomes. The current paper fills this gap in the literature and describes the implementation of \CRANpkg{bqror} package (version 1.6.0) for estimation and inference in Bayesian ordinal quantile regression.

The \CRANpkg{bqror} package offers two MCMC algorithms. Ordinal model with 3 or more outcomes is estimated through a combination of Gibbs sampling (Casella and George 1992) and Metropolis-Hastings (MH) algorithm (S. Chib and Greenberg 1995). The method is implemented in the function \texttt{quantregOR1}. For ordinal models with exactly 3 outcomes, the package presents a Gibbs sampling algorithm that is implemented in the function \texttt{quantregOR2}. We recommend using this procedure for an ordinal model with 3 outcomes, since its simpler and faster. Both functions, \texttt{quantregOR1} and \texttt{quantregOR2}, report the posterior mean, posterior standard deviation, 95\% posterior credible interval, and inefficiency factor of the model parameters. To compare alternative quantile regression models, we recommend using the marginal likelihood over the deviance information criterion (DIC). This is because the ``Bayesian approach'' to compare models is via the marginal likelihood (Siddhartha Chib 1995; Siddhartha Chib and Jeliazkov 2001). So, the \CRANpkg{bqror} package also provides the marginal likelihood with technical details for computation described in this paper. Additionally, the package includes functions to calculate the covariate effects and example codes to produce trace plots of MCMC draws. Lastly, this paper utilizes the \CRANpkg{bqror} package to demonstrate the estimation of quantile ordinal models on simulated data and real-life applications.

\hypertarget{sec:QROrdinal}{%
\section{Quantile regression in ordinal models}\label{sec:QROrdinal}}

Ordinal outcomes are common in a wide class of applications in economics, finance, marketing, social sciences, statistics in medicine, and transportation. In a typical study, the observed outcomes are ordered and categorical; so for the purpose of analysis scores/numbers are assigned to each outcome. For example, in a study on public opinion about offshore drilling (Mukherjee and Rahman 2016), responses may be recorded as follows: 1 for `strongly oppose', 2 for `somewhat oppose', 3 for `somewhat support', and 4 for `strongly support'. The numbers have an ordinal meaning but have no cardinal interpretation. We cannot interpret a score of 2 as twice the support compared to a score of 1, or the difference in support between 2 and 3 is the same as that between 3 and 4. With ordinal outcomes, the primary modeling objective is to express the probability of outcomes as a function of the covariates. Ordinal models that have been extensively studied and employed in applications include the ordinal probit and ordinal logit models (Johnson and Albert 2000; Greene and Hensher 2010), but they only give information about the average probability of outcomes conditional on the covariates.

Quantile regression with ordinal outcomes can be estimated using the monotone equivariance property and provides information on the probability of outcomes at different quantiles. In the spirit of Albert and Chib (1993), the ordinal quantile regression model can be presented in terms of an underlying latent (or unobserved) variable \(z_{i}\) as follows:
\begin{equation} 
z_{i} = x'_{i} \beta_{p} + \epsilon_{i}, \hspace{0.75in} \forall \; i=1, \cdots, n, 
\label{eq:latentreg} 
\end{equation}
where \(x'_{i}\) is a \(1 \times k\) vector of covariates, \(\beta_{p}\) is a \(k \times 1\) vector of unknown parameters at the \(p\)-th quantile, \(\epsilon_{i}\) follows an AL distribution i.e., \(\epsilon_{i} \sim AL(0,1,p)\), and \(n\) denotes the number of observations. Note that unlike the Classical (or Frequentist) quantile regression, the error is assumed to follow an AL distribution in order to construct a (working) likelihood (Yu and Moyeed 2001). The latent variable \(z_{i}\) is related to the observed discrete response \(y_{i}\) through the following relationship,
\begin{equation} 
\gamma_{p,j-1} < z_{i} \le \gamma_{p,j} \; \Rightarrow\; y_{i} = j, \hspace{0.75in} \forall \; i=1,\cdots, n; \; j=1,\cdots, J,
\label{eq:cutpoints} 
\end{equation}
where \(\gamma_{p} = (\gamma_{p,0}=-\infty, \gamma_{p,1},\ldots, \gamma_{p,J-1}, \gamma_{p,J}=\infty)\) is the cut-point vector and \(\textit{J}\) denotes the number of outcomes or categories. Typically, the cut-point \(\gamma_{p,1}\) is fixed at 0 to anchor the location of the distribution required for parameter identification (Jeliazkov and Rahman 2012). Given the observed data \(y\) = \((y_{1}, \cdots, y_{n})'\), the joint density (or likelihood when viewed as a function of the parameters) for the ordinal quantile model can be written as,
\begin{equation} 
f(y|\Theta_{p}) =  \prod_{i=1}^{n} \prod_{j=1}^{J} P(y_{i} = j | \Theta_{p})^{ I(y_{i} = j)}  
\label{eq:likelihood}
\end{equation}
where \(\Theta_{p} = (\beta_{p}, \gamma_{p})\), \(F_{AL}(\cdot)\) denotes the \({cdf}\) of an AL distribution and \(I(y_{i}=j)\) is an indicator function, which equals 1 if \(y_{i}=j\) and 0 otherwise.

Working directly with the AL likelihood \eqref{eq:likelihood} is not convenient for MCMC sampling. Therefore, the latent formulation of the ordinal quantile model \eqref{eq:latentreg}, following Kozumi and Kobayashi (2011), is expressed in the normal-exponential mixture form as follows,
\begin{equation}
z_{i} = x'_{i} \beta_{p} + \theta w_{i} + \tau \sqrt{w_{i}} \,u_{i}, \hspace{0.5in} \forall \; i=1, \cdots, n, \label{eq:latentregNE}
\end{equation}
where \(\epsilon_{i} = \theta w_{i} + \tau \sqrt{w_{i}} \, u_{i} \sim AL(0,1,p)\), \(w_{i} \sim\mathcal{E}(1)\) is mutually independent of \(u_{i} \sim N(0,1)\), \(N\) and \(\mathcal{E}\) denotes normal and exponential distributions, respectively; \(\theta = (1-2p)/[p(1-p)]\) and \(\tau = \sqrt{2/[p(1-p)]}\). Based on this formulation, we can write the conditional distribution of the latent variable as \(z_{i}|\beta_{p},w_{i} \sim N( x'_{i}\beta_{p} + \theta w_{i}, \tau^{2} w_{i})\) for \(i=1,\ldots,n\). This allows access to the properties of normal distribution which helps in constructing efficient MCMC algorithms.

\hypertarget{sec:ORI}{%
\subsubsection{\texorpdfstring{\(\mathrm{OR_{I}}\) model}{\textbackslash mathrm\{OR\_\{I\}\} model}}\label{sec:ORI}}

The term ``\(\mathrm{OR_{I}}\) model'' is ascribed to an ordinal model in which the number of outcomes (\(J\)) is equal to or greater than 3, location restriction is imposed by setting \(\gamma_{p,1} = 0\), and scale restriction is achieved via constant variance (See Rahman (2016); for a given value of \(p\), variance of a standard AL distribution is constant). Note that in contrast to Rahman (2016), our definition of \(OR_{I}\) model includes an ordinal model with exactly 3 outcomes. The location and scale restrictions are necessary to uniquely identify the parameters (see Jeliazkov, Graves, and Kutzbach (2008) and Jeliazkov and Rahman (2012) for further details and a pictorial representation).

During the MCMC sampling of the \(\mathrm{OR_{I}}\) model, we need to preserve the ordering of cut-points (\(\gamma_{p,0}=-\infty < \gamma_{p,1} < \gamma_{p,2} < \ldots < \gamma_{p,J-1} < \gamma_{p,J}=\infty\)). This is achieved by using a monotone transformation from a compact set to the real line. Many such transformations are available (e.g., log-ratios of category bin widths, arctan, arcsin), but the \CRANpkg{bqror} package utilizes the logarithmic transformation (Albert and Chib 2001; Rahman 2016),
\begin{equation} 
\delta_{p,j} = \ln ( \gamma_{p,j} - \gamma_{p,j-1} ), \qquad 2 \le j \le J-1. 
\label{eq:logtransformation}
\end{equation}
The cut-points \((\gamma_{p,1}, \gamma_{p,2}, \cdots, \gamma_{p,J-1})\) can be obtained from a one-to-one mapping to \((\delta_{p,2}, \cdots, \delta_{p,J-1})\).

With all the modeling ingredients in place, we employ the Bayes' theorem and express the joint posterior distribution as proportional to the product of the likelihood and the prior distributions. As in Rahman (2016), we employ independent normal priors: \(\beta_{p} \sim N(\beta_{p0}, B_{p0})\), \(\delta_{p} \sim N(\delta_{p0}, D_{p0})\) in the \CRANpkg{bqror} package. The augmented joint posterior distribution for the \(\mathrm{OR_{I}}\) model can thus be written as,
\begin{equation} 
\begin{split} 
& \pi(z,\beta_{p}, \delta_{p},w|y)  \propto f(y|z,\beta_{p},\delta_{p},w) \; \pi(z | \beta_{p}, w) \; \pi(w) \; \pi(\beta_{p}) \; \pi(\delta_{p}), \\
   & \propto \Big\{ \prod_{i=1}^{n} f(y_{i}|z_{i},\delta_{p}) \Big\} \; \pi(z | \beta_{p},w) \; \pi(w) \; \pi(\beta_{p}) \; \pi(\delta_{p}), \\
    & \propto  \prod_{i=1}^{n} \bigg\{ \prod_{j=1}^{J} 1\{\gamma_{p,j-1} < z_{i} \le \gamma_{p,j} \} \; N(z_{i}|x'_{i}\beta_{p} + \theta w_{i}, \tau^{2} w_{i}) \; \mathcal{E}(w_{i}|1) \bigg\} \\
    & \qquad \times \; N(\beta_{p}|\beta_{p0}, B_{p0}) \; N(\delta_{p}|\delta_{p0}, D_{p0}).
\end{split}
\label{eq:JointPostORI}
\end{equation}
where in the likelihood function of the second line, we use the fact that the observed \(y_{i}\) is independent of \((\beta_{p},w)\) given \((z_{i},\delta_{p})\). This follows from equation \eqref{eq:cutpoints} which shows that \(y_{i}\) given \((z_{i}, \delta_{p})\) is determined with probability 1. In the third line, we specify the conditional distribution of the latent variable and the prior distribution on the parameters.

The conditional posterior distributions are derived from the augmented joint posterior distribution \eqref{eq:JointPostORI}, and the parameters are sampled as per Algorithm\textasciitilde1. This algorithm is implemented in the \CRANpkg{bqror} package. The parameter \(\beta_{p}\) is sampled from an updated multivariate normal distribution and the latent weight \(w\) is sampled element-wise from a generalized inverse Gaussian (GIG) distribution. The cut-point vector \(\delta_{p}\) is sampled marginally of \((z, w)\) using a random-walk MH algorithm. Lastly, the latent variable \(z\) is sampled element-wise from a truncated normal (TN) distribution.

\vspace{6pt}

\(\textbf{Algorithm~1}\): Sampling in \(\mathrm{OR_{I}}\) model.

\noindent

\rule{\textwidth}{0.5pt}

\begin{itemize}
\tightlist
\item
  Sample \(\beta_{p}| z,w\) \(\sim\) \(N(\tilde{\beta}_{p}, \tilde{B}_{p})\), where,

  \begin{itemize}
  \tightlist
  \item
    \(\tilde{B}^{-1}_{p} = \bigg(\sum_{i=1}^{n} \frac{x_{i} x'_{i}}{\tau^{2} w_{i}} + B_{p0}^{-1} \bigg)\) \hspace{0.05in} and \hspace{0.05in} \(\tilde{\beta}_{p} = \tilde{B}_{p} \bigg(\sum_{i=1}^{n} \frac{x_{i}(z_{i} - \theta w_{i})}{\tau^{2} w_{i}} + B_{p0}^{-1} \beta_{p0} \bigg)\).
  \end{itemize}
\item
  Sample \(w_{i}|\beta_{p}, z_{i}\) \(\sim\) \(GIG\, (0.5, \tilde{\lambda}_{i}, \tilde{\eta})\), for \(i=1,\cdots,n\), where,

  \begin{itemize}
  \tightlist
  \item
    \(\tilde{\lambda}_{i} = \bigg(\frac{z_{i} - x'_{i}\beta_{p}}{\tau} \bigg)^{2}\) \hspace{0.05in} and \hspace{0.05in} \(\tilde{\eta} = \Big(\frac{\theta^{2}}{\tau^{2}} + 2 \Big)\).
  \end{itemize}
\item
  Sample \(\delta_{p}|y, \beta\) marginally of \(w\) (latent weight) and \(z\) (latent data), by generating \(\delta_{p}'\) using a random-walk chain \(\delta'_{p} = \delta_{p} + u\), where \(u \sim N(0_{J-2},\iota^{2} \hat{D})\), \(\iota\) is a tuning parameter and \(\hat{D}\) denotes the negative inverse Hessian, obtained by maximizing the log-likelihood with respect to \(\delta_{p}\). Given the current value of \(\delta_{p}\) and the proposed draw \(\delta'_{p}\), return \(\delta'_{p}\) with probability,
  \begin{equation*}
      \alpha_{MH}(\delta_{p}, \delta'_{p}) = \min \bigg\{1,\frac{ \;f(y|\beta_{p},\delta'_{p}) \;\pi(\beta_{p}, \delta'_{p})}{f(y|\beta_{p},\delta_{p}) \;\pi(\beta_{p}, \delta_{p})}\bigg\};
  \end{equation*}
  otherwise repeat the old value \(\delta_{p}\). The variance of \(u\) may be tuned as needed for appropriate step size and acceptance rate.
\item
  Sample \(z_{i}|y, \beta_{p}, \gamma_{p},w\) \(\sim\) \(TN_{(\gamma_{p, j-1}, \gamma_{p, j})}(x'_{i}\beta_{p} + \theta w_{i}, \tau^{2}w_{i})\) for \(i=1,\cdots,n\), where \(TN\) denotes a truncated normal distribution and \(\gamma_{p}\) is obtained via \(\delta_{p}\) using equation \eqref{eq:logtransformation}
\end{itemize}

\noindent

\rule{\textwidth}{0.5pt}

\hypertarget{sec:ORII}{%
\subsubsection{\texorpdfstring{\(\mathrm{OR_{II}}\) model}{\textbackslash mathrm\{OR\_\{II\}\} model}}\label{sec:ORII}}

The term ``\(\mathrm{OR_{II}}\) model'' is used for an ordinal model with exactly 3 outcomes (i.e., \(J = 3\)) where both location and scale restrictions are imposed by fixing the cut-points. Since there are only 2 cut-points and both are fixed at some values, the scale of the distribution needs to be free. Therefore, a scale parameter \(\sigma_{p}\) is introduced and the quantile ordinal model is rewritten as follows:
\begin{equation} 
\begin{split} 
 & z_{i} = x'_{i} \beta_{p} + \sigma_{p} \epsilon_{i}
          = x'_{i} \beta_{p}  + \sigma_{p} \theta w_{i} + \sigma_{p} \tau \sqrt{w_{i}} \,u_{i}, \hspace{0.8in} \forall \; i=1, \cdots, n, \\
& \gamma_{j-1} < z_{i} \le \gamma_{j} \; \Rightarrow \; y_{i} = j,
\hspace{1.6 in} \forall \; i=1,\cdots, n; \; j=1,2, 3, \end{split}
\label{eq:ORII}
\end{equation}
where \(\sigma_{p} \, \epsilon_{i} \sim AL(0, \sigma_{p}, p)\), (\(\gamma_{1},\gamma_{2}\)) are fixed, and recall \(\gamma_{0}=-\infty\) and \(\gamma_{3}=\infty\) . In this formulation, the conditional mean of \(z_{i}\) is dependent on \(\sigma_{p}\) which is problematic for Gibbs sampling. So, we define a new variable \(\nu_{i} = \sigma_{p} w_{i} \sim \mathcal{E}(\sigma_{p})\) and rewrite the model in terms of \(\nu_{i}\). In this representation, \(z_{i}|\beta_{p},\sigma_{p},\nu_{i} \sim N(x'_{i}\beta_{p} + \theta \nu_{i}, \tau^{2} \sigma_{p} \nu_{i})\), the conditional mean is free of \(\sigma_{p}\) and the model is conducive to Gibbs sampling.

The next step is to specify the prior distributions required for Bayesian inference. We follow Rahman (2016) and assume \(\beta_{p} \sim N(\beta_{p0}, B_{p0})\) and \(\sigma_{p} \sim IG(n_{0}/2, d_{0}/2)\); where \(IG\) stands for an inverse-gamma distribution. These are the default prior distributions in the \CRANpkg{bqror} package. Employing the Bayes' theorem, the augmented joint posterior distribution can be expressed as,
\begin{equation} 
\begin{split} 
& \pi(z,\beta_{p}, \nu, \sigma_{p} | y)
    \propto f(y|z, \beta_{p}, \nu, \sigma_{p} ) \; \pi(z|\beta_{p}, \nu, \sigma_{p} ) \; \pi(\nu|\sigma_{p}) \; \pi(\beta_{p}) \; \pi(\sigma_{p}),  \\
    & \propto \Big\{ \prod_{i=1}^{n} f(y_{i}|z_{i}, \sigma_{p})  \Big\} \; \pi(z|\beta_{p}, \nu, \sigma_{p} ) \; \pi(\nu|\sigma_{p}) \; \pi(\beta_{p}) \; \pi(\sigma_{p}), \\
    &  \propto \bigg\{ \prod_{i=1}^{n} \prod_{j=1}^{3} 1(\gamma_{j-1} < z_{i} \le \gamma_{j}) \; N(z_{i}|x'_{i}\beta_{p} + \theta \nu_{i}, \tau^{2} \sigma_{p} \nu_{i}) \; \mathcal{E}(\nu_{i}|\sigma_{p}) \bigg\} \\
    &   \qquad \times \; N(\beta_{p}|\beta_{p0}, B_{p0}) \; IG(\sigma_{p}|n_{0}/2, d_{0}/2),
\end{split} 
\label{eq:JointPostORII} 
\end{equation}
where the derivations in each step are analogous to those for the \(OR_{I}\) model.

The augmented joint posterior distribution, given by equation \eqref{eq:JointPostORII}, can be utilized to derive the conditional posterior distributions and the parameters are sampled as presented in Algorithm 2. This involves sampling \(\beta_{p}\) from an updated multivariate normal distribution and sampling \(\sigma_{p}\) from an updated IG distribution. The latent weight \(\nu\) is sampled element-wise from a GIG distribution and similarly, the latent variable \(z\) is sampled element-wise from a truncated normal distribution.

\vspace{6pt}

\(\textbf{Algorithm~2}\): Sampling in \(\mathrm{OR_{II}}\) model.

\noindent

\rule{\textwidth}{0.5pt}

\begin{itemize}
\tightlist
\item
  Sample \(\beta_{p}| z, \sigma_{p}, \nu\) \(\sim\) \(N(\tilde{\beta}_{p}, \tilde{B}_{p})\), where,

  \begin{itemize}
  \tightlist
  \item
    \(\tilde{B}^{-1}_{p} = \bigg(\sum_{i=1}^{n} \frac{x_{i}x'_{i}}{\tau^{2} \sigma_{p} \nu_{i}} + B_{p0}^{-1} \bigg)\) \hspace{0.05in} and \hspace{0.05in} \(\tilde{\beta}_{p} = \tilde{B}_{p}\bigg( \sum_{i=1}^{n} \frac{x_{i}(z_{i} - \theta\nu_{i})}{\tau^{2} \sigma_{p} \nu_{i}} + B_{p0}^{-1} \beta_{p0}\bigg)\)
  \end{itemize}
\item
  Sample \(\sigma_{p}| z, \beta_{p}, \nu\) \(\sim\) \(IG(\tilde{n}/2,\tilde{d}/2)\), where,

  \begin{itemize}
  \tightlist
  \item
    \(\tilde{n} = (n_{0} + 3n)\) \hspace{0.05in} and \hspace{0.05in} \(\tilde{d} = \sum_{i=1}^{n}(z_{i} - x'_{i}\beta_{p} - \theta\nu_{i})^{2}/\tau^{2}\nu_{i} + d_{0} + 2 \sum_{i=1}^{n} \nu_{i}\).
  \end{itemize}
\item
  Sample \(\nu_{i}| z_{i}, \beta_{p}, \sigma_{p}\) \(\sim\) \(GIG(0.5, \tilde{\lambda_{i}}, \tilde{\eta})\), for \(i=1,\cdots,n\), where,

  \begin{itemize}
  \tightlist
  \item
    \(\tilde{\lambda_{i}} = \frac{( z_{i} - x'_{i}\beta_{p})^2}{\tau^{2}\sigma_{p}}\) \hspace{0.05in} and \hspace{0.05in} \(\tilde{\eta} = \Big(\frac{\theta^2}{\tau^{2} \sigma_{p}} + \frac{2}{\sigma_{p}} \Big)\)
  \end{itemize}
\item
  Sample \(z_{i}|y, \beta_{p}, \sigma_{p}, \nu_{i}\) \(\sim\) \(TN_{(\gamma_{j-1}, \gamma_{j})}(x'_{i}\beta_{p} + \theta \nu_{i}, \tau^{2} \sigma_{p} \nu_{i})\) for \(i=1,\cdots,n\), and \(j=1,2,3\).
\end{itemize}

\noindent

\rule{\textwidth}{0.5pt}

\hypertarget{sec:ML}{%
\section{Marginal likelihood}\label{sec:ML}}

The article by Rahman (2016) employed DIC (Spiegelhalter et al. 2002; Gelman et al. 2013) for model comparison. However, in the Bayesian framework alternative models are typically compared using the marginal likelihood or the Bayes factor (Poirier 1995; Greenberg 2012). Therefore, we prefer using the marginal likelihood (or the Bayes factor) for comparing two or more regression models at a given quantile.

Consider a model \(\mathcal{M}_{s}\) with parameter vector \(\Theta_{s}\). Let \(f(y|\mathcal{M}_{s}, \Theta_{s})\) be its sampling density and \(\pi(\Theta_{s}|\mathcal{M}_{s})\) be the prior distribution; where \(s=1,\ldots,S\). Then, the marginal likelihood for the model \(\mathcal{M}_{s}\) is given by the expression, \(m(y|\mathcal{M}_{s}) = \int f(y|\Theta_{s},\mathcal{M}_{s}) \pi(\Theta_{s}|\mathcal{M}_{s}) \, d\Theta_{s}\). The Bayes factor is the ratio of marginal likelihoods. So, for any two models \(\mathcal{M}_{q}\) versus \(\mathcal{M}_{r}\), the Bayes factor, \(B_{qr} = \frac{m(y|\mathcal{M}_{q})}{m(y|\mathcal{M}_{r})} = \frac{ \int f(y| \mathcal{M}_{q}, \Theta_{q}) \; \pi(\Theta_{q}|\mathcal{M}_{q}) \; d\Theta_{q}} {\int f(y| \mathcal{M}_{r}, \Theta_{r}) \; \pi(\Theta_{r}|\mathcal{M}_{r}) \; d\Theta_{r}}\), can be easily computed once we have the marginal likelihoods.

Siddhartha Chib (1995) and later Siddhartha Chib and Jeliazkov (2001) showed that a simple and stable estimate of marginal likelihood can be obtained from the MCMC outputs. The approach is based on the recognition that the marginal likelihood can be written as the product of likelihood function and prior density over the posterior density. So, the marginal likelihood \(m(y|\mathcal{M}_{s})\) for model \(\mathcal{M}_{s}\) is expressed as,
\begin{equation}
m(y|\mathcal{M}_{s}) = \frac{f(y|\mathcal{M}_{s},\Theta_{s}) \, \pi(\Theta_{s}|\mathcal{M}_{s})}{\pi(\Theta_{s}|\mathcal{M}_{s},y)}.
\label{eq:BMLI} 
\end{equation}

Siddhartha Chib (1995) refers to equation \eqref{eq:BMLI} as the \(\textit{basic marginal likelihood identity}\) since it holds for all values in the parameter space, but typically computed at a high density point (e.g., mean, mode) denoted \(\Theta^{\ast}\) to minimize estimation variability. The likelihood ordinate \(f(y|\mathcal{M}_{s},\Theta^{\ast})\) is directly available from the model and the prior density \(\pi(\Theta^{\ast}|\mathcal{M}_{s})\) is assumed by the researcher. The novel part is the computation of posterior ordinate \(\pi(\Theta^{\ast}|y, \mathcal{M}_{s})\), which is estimated using the MCMC outputs. Since the marginal likelihood is often a large number, it is convenient to express it on the logarithmic scale. An estimate of the logarithm of marginal likelihood is given by,
\begin{equation} 
\ln \hat{m}(y) = \ln f(y|\Theta^{\ast}) + \ln \pi(\Theta^{\ast}) - \ln \hat{\pi}(\Theta^{\ast}|y), \label{eq:logBMLI}
\end{equation}
where we have dropped the conditioning on \(\mathcal{M}_{s}\) for notational simplicity. The next two subsections explain the computation of marginal likelihood for the \(OR_{I}\) and \(OR_{II}\) quantile regression models.

\hypertarget{marginal-likelihood-for-or_i-model}{%
\subsubsection{\texorpdfstring{Marginal likelihood for \(OR_{I}\) model}{Marginal likelihood for OR\_\{I\} model}}\label{marginal-likelihood-for-or_i-model}}

We know from \protect\hyperlink{sec:ORI}{Section 2.1} that the MCMC algorithm for estimating the \(OR_{I}\) model is defined by the following conditional posterior densities: \(\pi(\beta_{p}|z,w)\), \(\pi(\delta_{p}|\beta_{p},y)\), \(\pi(w|\beta_{p},z)\), and \(\pi(z|\beta_{p},\delta_{p},w,y)\). The conditional posteriors for \(\beta_{p}\), \(w\), and \(z\) have a known form, but that of \(\delta_{p}\) is not tractable and is sampled using an MH algorithm. Consequently, we adopt the approach of Siddhartha Chib and Jeliazkov (2001) to calculate the marginal likelihood for the \(OR_{I}\) model.

To simplify the computational process (specifically, to keep the computation over a reasonable dimension), we estimate the marginal likelihood marginally of the latent variables \((w,z)\). Moreover, we decompose the posterior ordinate as,
\begin{equation*} 
\pi(\beta_{p}^{\ast}, \delta_{p}^{\ast}|y) = \pi(\delta_{p}^{\ast}|y) \pi(\beta_{p}^{\ast}| \delta_{p}^{\ast},y),
\end{equation*}
where \(\Theta^{\ast} = (\beta_{p}^{\ast}, \delta_{p}^{\ast})\) denotes a high density point. By placing the intractable posterior ordinate first, we avoid the MH step in the \(\textit{reduced run}\) -- the process of running an MCMC sampler with one or more parameters fixed at some value -- of the MCMC sampler. We first estimate \(\pi(\delta_{p}^{\ast}|y)\) and then the reduced conditional posterior ordinate \(\pi(\beta_{p}^{\ast}| \delta_{p}^{\ast},y)\).

To obtain an estimate of \(\pi(\delta_{p}^{\ast}|y)\), we need to express it in a computationally convenient form. The parameter \(\delta_{p}\) is sampled using an MH step, which requires specifying a proposal density. Let \(q(\delta_{p}, \delta_{p}'|\beta_{p},w,z,y)\) denote the proposal density for the transition from \(\delta_{p}\) to \(\delta_{p}'\), and let,
\begin{equation} 
\alpha_{MH}(\delta_{p}, \delta'_{p}) = \min \bigg\{1, \frac{ \;f(y|\beta_{p},\delta'_{p}) \;\pi(\beta_{p}) \pi(\delta'_{p})} {f(y|\beta_{p},\delta_{p}) \;\pi(\beta_{p}) \pi(\delta_{p})} \times \frac{q(\delta'_{p}, \delta_{p}|\beta_{p},w,z,y)}{q(\delta_{p}, \delta'_{p} |\beta_{p},w,z,y)} \bigg\},
\label{eq:alphaMH} 
\end{equation}
denote the probability of making the move. In the context of the model, \(f(y|\beta_{p},\delta_{p})\) is the likelihood given by equation \eqref{eq:likelihood}, \(\pi(\beta_{p})\) and \(\pi(\delta_{p})\) are normal prior distributions (i.e., \(\beta_{p} \sim N(\beta_{p0}, B_{p0})\) and \(\delta_{p}\sim N(\delta_{p0}, D_{p0})\) as specified in \protect\hyperlink{sec:ORI}{Section 2.1}, and the proposal density \(q(\delta_{p}, \delta'_{p} |\beta_{p},w,z,y)\) is normal given by \(f_{N}(\delta'_{p}|\delta_{p}, \iota^{2}\hat{D})\) (see Algorithm\textasciitilde1 in \protect\hyperlink{sec:ORI}{Section 2.1}). There are two points to be noted about the proposal density. First, the conditioning on \((\beta_{p},w,z,y)\) is only for generality and not necessary as illustrated by the use of a random-walk proposal density. Second, since our MCMC sampler utilizes a random-walk proposal density, the second ratio on the right hand side of equation \eqref{eq:alphaMH} reduces to 1.

We closely follow the derivation in Siddhartha Chib and Jeliazkov (2001) and arrive at the following expression of the posterior ordinate,
\begin{equation} 
\pi(\delta_{p}^{\ast}|y) = \frac{E_{1} \{ \alpha_{MH}(\delta_{p},\delta_{p}^{\ast}|\beta_{p},w,z,y) \, q(\delta_{p}, \delta_{p}^{\ast}|\beta_{p},w,z,y) \} }{E_{2}\{ \alpha_{MH}(\delta_{p}^{\ast}, \delta_{p}|\beta_{p},w,z,y)\} },
\label{eq:ORIdeltaPostOrd} 
\end{equation}
where \(E_{1}\) represents expectation with respect to the distribution \(\pi(\beta_{p},\delta_{p},w,z|y)\) and \(E_{2}\) represents expectation with respect to the distribution \(\pi(\beta_{p},w,z|\delta_{p}^{\ast},y) \times q(\delta_{p}^{\ast}, \delta_{p}|\beta_{p},w,z,y)\). The quantities in equation \eqref{eq:ORIdeltaPostOrd} can be estimated using MCMC techniques. To estimate the numerator, we take the draw \(\{\beta_{p}^{(m)},\delta_{p}^{(m)},w^{(m)},z^{(m)}\}_{m=1}^{M}\) from the \(\textit{complete MCMC run}\) and take an average of the quantity \(\alpha_{MH}(\delta_{p}, \delta_{p}^{\ast}|\beta_{p},w,z,y) \, q(\delta_{p}, \delta_{p}^{\ast}|\beta_{p},w,z,y)\), where \(\alpha_{MH}(\cdot)\) is given by equation \eqref{eq:alphaMH} with \(\delta'_{p}\) replaced by \(\delta_{p}^{\ast}\), and \(q(\delta_{p}, \delta_{p}^{\ast}|\beta_{p},w,z,y)\) is given by the normal density \(f_{N}(\delta_{p}^{\ast}|\delta_{p}, \iota^{2} \hat{D})\).

The estimation of the quantity in the denominator is tricky! This requires generating an additional sample (say of \(H\) iterations) from the reduced conditional densities: \(\pi(\beta_{p}|w,z)\), \(\pi(w|\beta_{p},z)\), and \(\pi(z|\beta_{p},\delta_{p}^{\ast},w,y)\), where note that \(\delta_{p}\) is fixed at \(\delta_{p}^{\ast}\) in the MCMC sampling, and thus corresponds to a \(\textit{reduced run}\). Moreover, at each iteration, we generate
\begin{equation*} 
\delta_{p}^{(h)} \sim q(\delta_{p}^{\ast},\delta_{p}|\beta_{p}^{(h)},w^{(h)},z^{(h)},y) \equiv f_{N}(\delta_{p}|\delta_{p}^{\ast}, \iota^{2} \hat{D}).
\end{equation*}
The resulting quadruplet of draws \(\{\beta_{p}^{(h)}, \delta_{p}^{(h)}, w^{(h)}, z^{(h)} \}\), as required, is a sample from the distribution \(\pi(\beta_{p},w,z|\delta_{p}^{\ast},y) \times q(\delta_{p}^{\ast}, \delta_{p}|\beta_{p},w,z,y)\). With the numerator and denominator now available, the posterior ordinate \(\pi(\delta_{p}^{\ast}|y)\) is estimated as,

\begin{equation}
\hat{\pi}(\delta_{p}^{\ast}|y) = \frac{ M^{-1} \sum_{m=1}^{M} \{ \alpha_{MH} (\delta_{p}^{(m)}, \delta_{p}^{\ast}|\Lambda^{(m)},y) \, q(\delta_{p}^{(m)}, \delta_{p}^{\ast}|\Lambda^{(m)},y) \} } {H^{-1} \sum_{h=1}^{H} \{ \alpha_{MH}(\delta_{p}^{\ast}, \delta_{p}^{(h)}|\Lambda^{(h)},y)\} }.
\label{eq:ORIdeltaPostOrdEstimate} 
\end{equation}
where \(\Lambda^{(m)} = (\beta_{p}^{(m)},w^{(m)},z^{(m)})\) and \(\Lambda^{(h)}=(\beta_{p}^{(h)},w^{(h)},z^{(h)})\).

The computation of the posterior ordinate \(\pi(\beta_{p}^{\ast}| \delta_{p}^{\ast},y)\) is trivial. We have the sample of \(H\) draws \(\{ w^{(h)}, z^{(h)} \}\) from the reduced run, which are marginally of \(\beta_{p}\) from the distribution \(\pi(w,z|\delta_{p}^{\ast},y)\). These draws are utilized to estimate the posterior ordinate as,
\begin{equation} 
\hat{\pi}(\beta_{p}^{\ast}|\delta_{p}^{\ast},y) = H^{-1} \sum_{h=1}^{H} \pi(\beta_{p}^{\ast}|\delta_{p}^{\ast},w^{(h)},z^{(h)},y).
\label{eq:ORIbetaPostOrdEstimate} 
\end{equation}
Substituting the two density estimates given by equations \eqref{eq:ORIdeltaPostOrdEstimate} and \eqref{eq:ORIbetaPostOrdEstimate} into equation \eqref{eq:logBMLI}, an estimate of the logarithm of marginal likelihood for \(OR_{I}\) model is obtained as,
\begin{equation} 
\ln \hat{m}(y) = \ln f(y|\beta_{p}^{\ast},\delta_{p}^{\ast}) + \ln \Big[ \pi(\beta_{p}^{\ast}) \pi(\delta_{p}^{\ast}) \Big] - \ln \Big[ \hat{\pi}(\delta_{p}^{\ast}|y) \, \hat{\pi}(\beta_{p}^{\ast}|\delta_{p}^{\ast},y) \Big], 
\label{eq:ORIMLestimate}
\end{equation}
where the likelihood \(f(y|\beta_{p}^{\ast},\delta_{p}^{\ast})\) and prior densities are evaluated at \(\Theta^{\ast} = (\beta_{p}^{\ast}, \delta_{p}^{\ast})\).

\hypertarget{marginal-likelihood-for-or_ii-model}{%
\subsubsection{\texorpdfstring{Marginal likelihood for \(OR_{II}\) model}{Marginal likelihood for OR\_\{II\} model}}\label{marginal-likelihood-for-or_ii-model}}

We know from \protect\hyperlink{sec:ORII}{Section 2.2} that the \(OR_{II}\) model is estimated by Gibbs sampling and hence we follow Siddhartha Chib (1995) to compute the marginal likelihood. The Gibbs sampler consists of four conditional posterior densities given by \(\pi(\beta_{p}|\sigma_{p},\nu,z)\), \(\pi(\sigma_{p}|\beta_{p},\nu,z)\), \(\pi(\nu|\beta_{p}, \sigma_{p},z)\), and \(\pi(z|\beta_{p},\sigma_{p},\nu,y)\). However, the variables \((\nu, z)\) are latent. So, we integrate them out and write the posterior ordinate as \(\pi(\beta_{p}^{\ast},\sigma_{p}^{\ast}|y) = \pi(\beta_{p}^{\ast}|y) \pi(\sigma_{p}^{\ast}|\beta_{p}^{\ast},y)\), where the terms on the right hand side can be written as,
\begin{equation*} 
\begin{split} 
\pi(\beta_{p}^{\ast}|y) & = \int \pi(\beta_{p}^{\ast}|\sigma_{p},\nu,z,y) \, \pi(\sigma_{p},\nu,z|y) \, d\sigma_{p} \, d\nu \, dz, \\ \pi(\sigma_{p}^{\ast}|\beta_{p}^{\ast},y) & = \int \pi(\sigma_{p}^{\ast}|\beta_{p}^{\ast},\nu,z,y) \,\pi(\nu,z|\beta_{p}^{\ast},y) \, d\nu \, dz, \end{split}
\end{equation*}
and \(\Theta^{\ast} = (\beta_{p}^{\ast}, \sigma_{p}^{\ast})\) denotes a high density point, such as the mean or the median.

The posterior ordinate \(\pi(\beta_{p}^{\ast}|y)\) can be estimated as the ergodic average of the conditional posterior density with the posterior draws of \((\sigma_{p},\nu,z)\). Therefore, \(\pi(\beta_{p}^{\ast}|y)\) is estimated as,
\begin{equation} 
\hat{\pi}(\beta_{p}^{\ast}|y) = G^{-1} \sum_{g=1}^{G} \pi(\beta_{p}^{\ast}|\sigma_{p}^{(g)}, \nu^{(g)}, z^{(g)},y ).
\label{eq:ORIIbetaPostOrd} 
\end{equation}
The term \(\pi(\sigma_{p}^{\ast}|\beta_{p}^{\ast},y)\) is a reduced conditional density ordinate and can be estimated with the help of a \(\textit{reduce run}\). So, we generate an additional sample (say another \(G\) iterations) of \(\{ \nu^{(g)},z^{(g)} \}\) from \(\pi(\nu,z|\beta_{p}^{\ast},y)\) by successively sampling from \(\pi(\sigma_{p}|\beta_{p}^{\ast},\nu,z)\), \(\pi(\nu|\beta_{p}^{\ast}, \sigma_{p},z)\), and \(\pi(z|\beta_{p}^{\ast},\sigma_{p},\nu,y)\), where note that \(\beta_{p}\) is fixed at \(\beta_{p}^{\ast}\) in each conditional density. Next, we use the draws \(\{ \nu^{(g)},z^{(g)} \}\) to compute,
\begin{equation} 
\hat{\pi}(\sigma_{p}^{\ast}|\beta_{p}^{\ast},y) = G^{-1} \sum_{g=1}^{G} \pi(\sigma_{p}^{\ast}|\beta_{p}^{\ast},\nu^{(g)}, z^{(g)},y).
\label{eq:ORIIsigmaPostOrd}
\end{equation}
which is a simulation consistent estimate of \(\pi(\sigma_{p}^{\ast}|\beta_{p}^{\ast},y)\).

Substituting the two density estimates given by equations \eqref{eq:ORIIbetaPostOrd} and \eqref{eq:ORIIsigmaPostOrd} into equation \eqref{eq:logBMLI}, we obtain an estimate of the logarithm of marginal likelihood,
\begin{equation} 
\ln \hat{m}(y) = \ln f(y|\beta_{p}^{\ast},\sigma_{p}^{\ast}) + \ln \Big[ \pi(\beta_{p}^{\ast}) \pi(\sigma_{p}^{\ast}) \Big] - \ln \Big[ \hat{\pi}(\beta_{p}^{\ast}|y) \,
\hat{\pi}(\sigma_{p}^{\ast}|\beta_{p}^{\ast},y) \Big],
\label{eq:ORIIMLestimate}
\end{equation}
where the likelihood function and prior densities are evaluated at \(\Theta^{\ast} = (\beta_{p}^{\ast}, \sigma_{p}^{\ast})\). Here, the likelihood function has the expression,
\begin{equation*} 
f(y|\beta_{p}^{\ast},\sigma_{p}^{\ast}) =  \prod_{i=1}^{n}  \prod_{j=1}^{3} \bigg[F_{AL}\left(\frac{\gamma_{j} - x'_{i}\beta_{p}^{\ast}}{\sigma_{p}^{\ast}}\right) - F_{AL}\left(\frac{\gamma_{j-1} - x'_{i}\beta_{p}^{\ast}}{\sigma_{p}^{\ast}}\right) \bigg]^{ I(y_{i} = j)},
\end{equation*}
where the cut-points \(\gamma\) are known and fixed for identification reasons as explained in \protect\hyperlink{sec:ORII}{Section 2.2}.

\hypertarget{sec:SimStudies}{%
\section{Simulation studies}\label{sec:SimStudies}}

This section explains the data generating process for simulation studies, the functions offered in the \CRANpkg{bqror} package, and usage of the functions for estimation and inference in ordinal quantile models.

\hypertarget{subsec:SimDataORI}{%
\subsubsection{\texorpdfstring{\(OR_{I}\) model: data, functions, and outputs}{OR\_\{I\} model: data, functions, and outputs}}\label{subsec:SimDataORI}}

\(\textit{Data Generation}\): The data for simulation study in \(OR_{I}\) model is generated from the regression: \(z_{i} = x'_{i}\beta + \epsilon_{i}\), where \(\beta = (-4, 5, 6)\), \((x_{2},x_{3}) \sim U(0, 1)\), and \(\epsilon_{i} \sim AL(0, \sigma = 1, p)\) for \(i=1,\ldots,n\). Here, \(U\) and \(AL\) denote a uniform distribution and an asymmetric Laplace distribution, respectively. The \(z\) values are continuous and are classified into 4 categories based on the cut-points \((0, 2, 4)\) to generate ordinal values of \(y\), the outcome variable. We follow the above procedure to generate 3 data sets each with 500 observations (i.e., \(n = 500\)). The 3 data sets correspond to the quantile \(p\) equaling 0.25, 0.50, and 0.75, and are stored as \texttt{data25j4}, \texttt{data50j4}, and \texttt{data75j4}, respectively. Note that the last two letters in the name of the data (i.e., \texttt{j4}) denote the number of unique outcomes in the \(y\) variable.

We now describe the major functions for Bayesian quantile estimation of \(OR_{I}\) model, demonstrate their usage, and note the inputs and outputs of each function.

\(\textbf{quantregOR1}\): The \texttt{quantregOR1} is the primary function for estimating Bayesian quantile regression in ordinal models with 3 or more outcomes (i.e., \(OR_{I}\) model) and implements Algorithm\textasciitilde1. In the code snippet below, we first read the data and then do the following: define the ordinal response variable (\texttt{y}) and covariate matrix (\texttt{xMat}), specify the number of covariates (\texttt{k}) and number of outcomes (\texttt{J}), and set the prior means and covariances for \(\beta_{p}\) and \(\delta_{p}\). We then call the \texttt{quantregOR1} function and specify the inputs: ordinal outcome variable (\texttt{y}), covariate matrix including a column of ones (\texttt{xMat}), prior mean (\texttt{b0}) and prior covariance matrix (\texttt{B0}) for the regression coefficients, prior mean (\texttt{d0}) and prior covariance matrix (\texttt{D0}) for the transformed cut-points, burn-in size (\texttt{burn}), post burn-in size (\texttt{mcmc}), quantile (\texttt{p}), the tuning factor (\texttt{tune}) to adjust the MH acceptance rate, and the auto correlation cutoff value (\texttt{accutoff}). The last input \texttt{verbose}, when set to \texttt{TRUE} (\texttt{FALSE}) will (not) print the summary output.

In the code below, we use a diffuse normal prior \(\beta_{p} \sim N(0_{k}, 10\ast I_{k})\) where \(0_{k}\) and \(I_{k}\) are matrices of dimension \(k \times 1\) and \(k \times k\), respectively. The prior distribution can be further diffused (i.e., made less informative) by increasing 10 to say 100. Besides, the prior variance for \(\delta_{p}\) should be small, such as \(0.25 \ast I_{J-2}\), since the distribution is on the transformed cut-points, which is in the logarithmic scale. If there is a need for prior elicitation, they can be designed from previous subject based knowledge or by estimating the model on a training sample and then using the results to form prior distributions (See Greenberg (2012), for examples.)

\begin{verbatim}
library('bqror')
data("data25j4")
y <- data25j4$y
xMat <- data25j4$x
k <- dim(xMat)[2]
J <- dim(as.array(unique(y)))[1]
b0 <- array(rep(0, k), dim = c(k, 1))
B0 <- 10*diag(k)
d0 <- array(0, dim = c(J-2, 1))
D0 <- 0.25*diag(J - 2)
modelORI <- quantregOR1(y = y, x = xMat, b0, B0, d0, D0, burn = 1125,
                        mcmc = 4500, p = 0.25, tune = 1, accutoff = 0.5,
                        verbose = TRUE)

[1] Summary of MCMC draws:

        Post Mean Post Std Upper Credible Lower Credible  Inef Factor
beta_1    -3.6434   0.4293        -2.8369        -4.5073       2.3272
beta_2     4.8283   0.5597         5.9577         3.7624       2.4529
beta_3     5.9929   0.5996         7.2474         4.8819       2.7491
delta_1    0.7152   0.1110         0.9616         0.5004       3.2261
delta_2    0.7456   0.0940         0.9281         0.5543       2.1497

[1] MH acceptance rate: 31.82%
[1] Log of Marginal Likelihood: -545.72
[1] DIC: 1060.56
\end{verbatim}

The outputs from the \texttt{quantregOR1} function are the following quantities: \texttt{summary.bqrorOR1}, \texttt{postMeanbeta}, \texttt{postMeandelta}, \texttt{postStdbeta}, \texttt{postStddelta}, \texttt{gammacp}, \texttt{acceptancerate}, \texttt{logMargLike}, \texttt{dicQuant}, \texttt{ineffactor}, \texttt{betadraws}, and \texttt{deltadraws}. A detailed description of each output is presented in the \CRANpkg{bqror} package help file. In the summary, we report the posterior mean, posterior standard deviation, 95\% posterior credible (or probability) interval, and the inefficiency factor of the quantile regression coefficients \(\beta_{p}\) and transformed cut-points \(\delta_{p}\). These quantities are presented in the last five columns and labeled appropriately.

The posterior means of (\(\beta_{p}, \delta_{p}\)) are close to the true values used to generate the data with small standard deviations. So, the \texttt{quantregOR1} function is successful in recovering the true values of the parameters. The inefficiency factor is computed from the MCMC samples using the batch-means method (Greenberg 2012). They indicate the cost of working with correlated samples. For example, an inefficiency factor of 3 implies that it takes 3 correlated draws to get one independent draw. As such, low inefficiency factor indicates better mixing and a more efficient MCMC algorithm. Inefficiency factor also bears a direct relationship with effective sample size, where the latter can be obtained as the total number of (post burn-in) MCMC draws divided by the inefficiency factor (Siddhartha Chib 2012). The inefficiency factors for (\(\beta_{p}\), \(\delta_{p}\)) are stored in the object \texttt{modelOR1} of class \texttt{bqrorOR1} and can be obtained by calling \texttt{modelORI\$ineffactor}.

The third last row displays the random-walk MH acceptance rate for \(\delta_{p}\), for which the preferred acceptance rate is around 30 percent. The last two rows present the model comparison measures, the logarithm of marginal likelihood and the DIC. The logarithm of marginal likelihood is computed using the MCMC outputs from the complete and reduced runs as explained in \protect\hyperlink{sec:ML}{Section 3}, while the principle for computing the DIC is presented in Gelman et al. (2013). For any two competing models at the same quantile, the model with a higher (lower) marginal likelihood (DIC) provides a better model fit.

While the two model comparison measures are printed as part of the summary output, they can also be called individually. For example, the logarithm of marginal likelihood can be obtained by calling \texttt{modelORI\$logMargLike}. Whereas, the DIC can be obtained by calling \texttt{modelORI\$dicQuant\$DIC}. Two more quantities that are part of the object \texttt{modelORI\$dicQuant} are effective number of parameters denoted \(p_{D}\) and the deviance computed at the posterior mean. They can be obtained by calling \texttt{modelORI\$dicQuant\$pd} and \texttt{modelORI\$dicQuant\$dev}, respectively. Besides, post estimation, one may also use the command \texttt{modelORI\$summary} or \texttt{summary.bqrorOR1(modelORI)} to extract and print the summary output.

\(\textbf{covEffectOR1}\): The function \texttt{covEffectOR1} computes the average covariate effect for different outcomes of \(OR_{I}\) model at a specified quantile, marginally of the parameters and the remaining covariates. While a demonstration of this function is best understood in a real-life study and is presented in the application section, here we present the mechanics behind the computation of average covariate effect.

Suppose, we want to compute the average covariate effect when the \(l\)-th covariate \(\{x_{i,l}\}\) is set to the values \(a\) and \(b\), denoted as \(\{x_{i,l}^{a}\}\) and \(\{ x_{i,l}^{b}\}\), respectively. We split the covariate and parameter vectors as follows: \(x_{i}^{a} = ( x_{i,l}^{a}, \, x_{i,-l})\), \(x_{i}^{b} = (x_{i,l}^{b}, \, x_{i,-l})\), and \(\beta_{p} = (\beta_{p,l}, \, \beta_{p,-l})\), where \(-l\) in the subscript denotes all covariates (parameters) except the \(l\)-th covariate (parameter). We are interested in the distribution of the difference \(\{\Pr(y_{i}=j|x_{i,l}^{b}) - \Pr(y_{i}=j|x_{i,l}^{a} ) \}\) for \(1 \le j \le J\), marginalized over \(\{x_{i,-l}\}\) and \((\beta_{p},\delta_{p})\), given the data \(y=(y_{1}, \cdots, y_{n})'\). We marginalize the covariates using their empirical distribution and the parameters based on the posterior distribution.

To obtain draws from the distribution \(\{\Pr(y_{i}=j|x_{i,l}^{b}) - \Pr(y_{i}=j|x_{i,l}^{a} ) \}\), we use the method of composition (See Siddhartha Chib and Jeliazkov (2006), Rahman and Vossmeyer (2019), and Bresson, Lacroix, and Rahman (2021), for additional details). In this process, we randomly select an individual, extract the corresponding sequence of covariate values, draw a value \((\beta_{p}, \delta_{p})\) from the posterior distributions, and evaluate \(\{\Pr(y_{i}=j|x_{i}^{b},\beta_{p},\delta_{p}) - \Pr(y_{i}=j|x_{i}^{a},\beta_{p},\delta_{p} )\}\), where,
\begin{equation*}
\begin{split}
& \Pr(y_{i}=j|x_{i}^{q},\beta_{p},\delta_{p}) \\
& = F_{AL}(\gamma_{p,j} - x_{i,l}^{q} \, \beta_{p,l} - x'_{i,-l} \, \beta_{p,-l}) - F_{AL}(\gamma_{p,j-1} - x_{i,l}^{q} \, \beta_{p,l} - x'_{i,-l} \, \beta_{p,-l}),
\end{split}
\end{equation*}
for \(q=b, a\) and \(1 \le j \le J\). This process is repeated for all remaining individuals and other MCMC draws from the posterior distribution. Finally, the average covariate effect \((ACE)\) for outcome \(j\) (for \(1 \le j \le J\)) is calculated as the mean of the difference in pointwise probabilities,
\begin{equation}
\frac{1}{M} \frac{1}{n} \sum_{m=1}^{M}
\sum_{i=1}^{n} \Big[ \Pr(y_{i}=j|x_{i}^{b},\beta_{p}^{(m)},\delta_{p}^{(m)}) - \Pr(y_{i}=j|x_{i}^{a},\beta_{p}^{(m)},\delta_{p}^{(m)} ) \Big],
\label{eq:ACEOR1}
\end{equation}
where \((\beta_{p}^{(m)},\delta_{p}^{(m)})\) is an MCMC draw of \((\beta_{p}, \delta_{p})\), and \(M\) is the number of post burn-in MCMC draws.

\hypertarget{subsec:SimDataORII}{%
\subsubsection{\texorpdfstring{\(OR_{II}\) model: data, function, and outputs}{OR\_\{II\} model: data, function, and outputs}}\label{subsec:SimDataORII}}

\(\textit{Data Generation}\): The data generating process for the \(OR_{II}\) model closely resembles that of \(OR_{I}\) model. In particular, 500 observations are generated for each value of \(p\) from the regression model: \(z_{i} = x'_{i}\beta + \epsilon_{i}\), where \(\beta = (-4, 6, 5)\), \((x_{2},x_{3}) \sim U(0, 1)\) and \(\epsilon_{i} \sim AL(0, \sigma = 1, p)\) for \(i=1,\ldots,n\). The continuous values of \(z\) are classified based on the cut-points \((0, 3)\) to generate 3 ordinal values for \(y\), the outcome variable. Once again, we choose \(p\) equal to 0.25, 0.50, and 0.75 to generate three samples from the model, which are referred to as \texttt{data25j3}, \texttt{data50j3}, and \texttt{data75j3}, respectively. The last two letters in the data names (i.e., \texttt{j3}) denote the number of unique outcomes in the \(y\) variable.

\(\textbf{quantregOR2}\): The function \texttt{quantregOR2} implements Algorithm\textasciitilde2 and is the main function and for estimating Bayesian quantile regression in \(OR_{II}\) model i.e., an ordinal model with exactly 3 outcomes. In the code snippet below, we first read the data, define the required quantities, and then call the \texttt{quantregOR2} for estimating the quantile model. The function inputs are as follows: the ordinal outcome variable (\texttt{y}), covariate matrix including a column of ones (\texttt{xMat}), prior mean (\texttt{b0}) and prior covariance matrix (\texttt{B0}) for \(\beta_{p}\), prior shape (\texttt{n0}) and scale (\texttt{d0}) parameters for \(\sigma_{p}\), second cut-point (\texttt{gammacp2}), burn-in size (\texttt{burn}), post burn-in size (\texttt{mcmc}), quantile (\texttt{p}), auto correlation cutoff value (\texttt{accutoff}), and the \texttt{verbose} option which when set to \texttt{TRUE} (\texttt{FALSE}) will (not) print the outputs. We use a relatively diffuse prior distributions on \((\beta_{p}, \sigma_{p})\) to allow the data to speak for itself.

\begin{verbatim}
library('bqror')
data("data25j3")
y <- data25j3$y
xMat <- data25j3$x
k <- dim(xMat)[2]
b0 <- array(rep(0, k), dim = c(k, 1))
B0 <- 10*diag(k)
n0 <- 5
d0 <- 8
modelORII <- quantregOR2(y = y, x = xMat, b0, B0, n0, d0, gammacp2 = 3,
                         burn = 1125, mcmc = 4500, p = 0.25, accutoff = 0.5,
                         verbose = TRUE)

[1] Summary of MCMC draws:

       Post Mean Post Std Upper Credible Lower Credible  Inef Factor
beta_1   -3.8900   0.4560        -3.0578        -4.8346       2.4784
beta_2    5.8257   0.5341         6.9263         4.8243       2.1231
beta_3    4.7194   0.5227         5.7502         3.7194       2.2693
sigma     0.8968   0.0763         1.0587         0.7626       2.4079

[1] Log of Marginal Likelihood: -404.34
[1] DIC: 790.72
\end{verbatim}

The outputs from the \texttt{quantregOR2} function are the following: \texttt{summary.bqrorOR2}, \texttt{postMeanbeta}, \texttt{postMeansigma}, \texttt{postStdbeta}, \texttt{postStdsigma}, \texttt{logMargLike}, \texttt{dicQuant}, \texttt{ineffactor}, \texttt{betadraws}, and \texttt{sigmadraws}. A detailed description of each output is presented in the \CRANpkg{bqror} package help file. Once again, we summarize the MCMC draws by reporting the posterior mean, posterior standard deviation, 95\% posterior credible (or probability) interval, and the inefficiency factor of the quantile regression coefficients \(\beta_{p}\) and scale parameter \(\sigma_{p}\). These quantities are presented in the last five columns and labeled appropriately.

The output also exhibits the logarithm of marginal likelihood and the DIC for \(OR_{II}\) model. For two or more models at the same quantile, the model with a higher (lower) marginal likelihood (DIC) indicates a better model fit. The logarithm of marginal likelihood is computed using the Gibbs output as explained in \protect\hyperlink{sec:ML}{Section 3}, while the computation of DIC follows Gelman et al. (2013).

The two model comparison measures can also be printed individually. For example, the logarithm of marginal likelihood can be obtained by calling \texttt{modelORII\$logMargLike} and the DIC by calling \texttt{modelORII\$dicQuant\$DIC}. The effective number of parameters \(p_{D}\) and the deviance computed at the posterior mean can obtained by calling \texttt{modelORII\$dicQuant\$pd} and \texttt{modelORII\$dicQuant\$dev}, respectively. Lastly, one may use the command \texttt{modelORII\$summary} or \texttt{summary.bqrorOR2(modelORII)} to extract and print the summary output.

\(\textbf{covEffectOR2}\): The function \texttt{covEffectOR2} computes the average covariate effect for the 3 outcomes of \(OR_{II}\) model at a specified quantile, marginally of the parameters and remaining covariates. The principle underlying the computation is analogous to that of \(OR_{I}\) model and explained below. An implementation of the function is presented in the tax policy application.

Suppose, we are interested in computing the average covariate effect for the \(l\)-th covariate \(\{x_{i,l}\}\) for two different values \(a\) and \(b\), and split the covariate and parameter vectors as: \(x_{i}^{a} = ( x_{i,l}^{a}, \,x_{i,-l})\), \(x_{i}^{b} = (x_{i,l}^{b}, \, x_{i,-l})\), \(\beta = (\beta_{p,l}, \, \beta_{p,-l})\). We are interested in the distribution of the difference \(\{\Pr(y_{i}=j|x_{i,l}^{b}) - \Pr(y_{i}=j|x_{i,l}^{a} ) \}\) for \(1 \le j \le J=3\), marginalized over \(\{x_{i,-l}\}\) and \((\beta_{p},\sigma_{p})\), given the data \(y=(y_{1}, \cdots, y_{n})'\). We again employ the method of composition i.e., randomly select an individual, extract the corresponding sequence of covariate values, draw a value \((\beta_{p}, \sigma_{p})\) from their posterior distributions, and lastly evaluate \(\{\Pr(y_{i}=j|x_{i}^{b},\beta_{p},\sigma_{p}) - \Pr(y_{i}=j|x_{i}^{a},\beta_{p},\sigma_{p} )\}\), where
\begin{equation*}
\begin{split}
& \Pr(y_{i}=j|x_{i}^{q},\beta_{p},\sigma_{p}) \\
& = F_{AL}\bigg( \frac{\gamma_{p,j} - x_{i,l}^{q} \, \beta_{p,l} - x'_{i,-l} \,          \beta_{p,-l}}{\sigma_{p}} \bigg) - F_{AL}\bigg( \frac{\gamma_{p,j-1} - x_{i,l}^{q} \, \beta_{p,l} - x'_{i,-l} \, \beta_{p,-l}}{\sigma_{p}} \bigg),
\end{split}
\end{equation*}
for \(q=b, a\), and \(1 \le j \le J=3\). This process is repeated for other individuals and the remaining Gibbs draws to compute the \(ACE\) for outcome \(j\) (\(=1,2,3\)) as below,
\begin{equation}
  \frac{1}{G} \frac{1}{n} \sum_{g=1}^{G}
    \sum_{i=1}^{n} \Big[ \Pr(y_{i}=j|x_{i}^{b},\beta_{p}^{(g)},\sigma_{p}^{(g)}) - \Pr(y_{i}=j|x_{i}^{a},\beta_{p}^{(g)},\sigma_{p}^{(g)} ) \Big],
    \label{eq:ACEOR2}
\end{equation}
where \((\beta_{p}^{(g)},\sigma_{p}^{(g)})\) is a Gibbs draw of \((\beta_{p}, \sigma_{p})\) and \(G\) is the number of post burn-in Gibbs draws.

\hypertarget{sec:Applications}{%
\section{Applications}\label{sec:Applications}}

In this section, we consider the educational attainment and tax policy applications from Rahman (2016) to demonstrate the real data applications of the proposed \CRANpkg{bqror} package. While the educational attainment study displays the implementation of ordinal quantile regression in \(OR_{I}\) model, the tax policy study highlights the use of ordinal quantile regression in \(OR_{II}\) model. Data for both the applications are included as a part of the \CRANpkg{bqror} package.

\hypertarget{subsec:EducAttainment}{%
\subsubsection{Educational attainment}\label{subsec:EducAttainment}}

\begin{figure}

{\centering \includegraphics[width=0.55\linewidth,height=0.24\textheight]{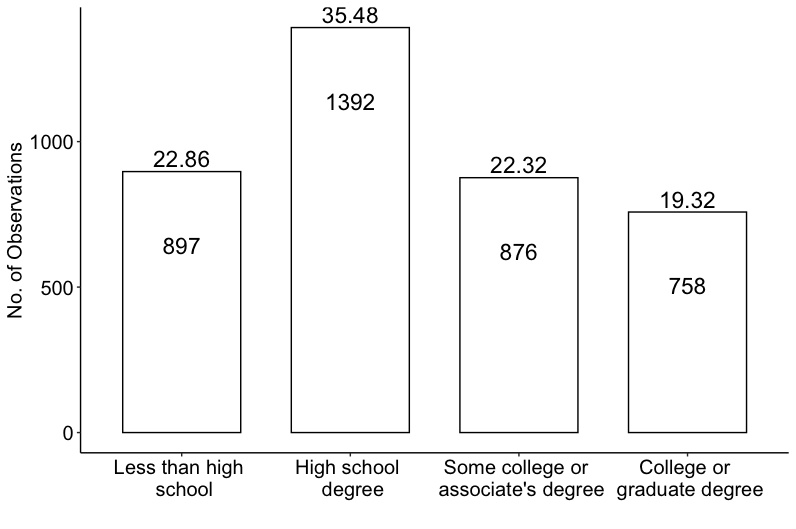} 

}

\caption{Bar chart showing the different categories of educational attainment. The number of responses (percentage) for each category are shown inside (at the top) of each bar.}\label{fig:EducFreqDist-tab-static}
\end{figure}

In this application, the goal is to study the effect of family background, individual level variables, and age cohort on educational attainment of 3923 individuals using data from the National Longitudinal Study of Youth (NLSY, 1979) (Jeliazkov, Graves, and Kutzbach 2008; Rahman 2016). The dependent variable in the model, education degrees, has four categories: \(\textit{(i) Less than high school}\), \(\textit{(ii) High school degree}\), \(\textit{(iii) Some college or associate's degree}\), and \(\textit{(iv) College or graduate degree}\). A bar chart of the four categories is presented in Figure \ref{fig:EducFreqDist-tab-static}. The independent variables in the model include intercept, square root of family income, mother's education, father's education, mother's working status, gender, race, indicator variables to point whether the youth lived in an urban area or South at the age of 14, and three indicator variables to indicate the individual's age in 1979 (serves as a control for age cohort effects).

To estimate the Bayesian ordinal quantile model on educational attainment data, we load the \CRANpkg{bqror} package, prepare the required inputs and feed them into the \texttt{quantregOR1} function. Specifically, we specify the outcome variable, covariate matrix (with covariates in order as in Rahman 2016), prior distributions for \((\beta_{p}, \delta_{p})\), burn-in size, number of post burn-in MCMC iterations, quantile value (\(p=0.5\) for this illustration), and the values for tuning factor and autocorrelation cutoff.

\begin{verbatim}
library('bqror')
data("Educational_Attainment")
data <- na.omit(Educational_Attainment)
data$fam_income_sqrt <- sqrt(data$fam_income)
cols <- c("mother_work","urban","south", "father_educ","mother_educ",
          "fam_income_sqrt","female", "black","age_cohort_2","age_cohort_3",
          "age_cohort_4")
x <- data[cols]
x$intercept <- 1
xMat <- x[,c(12,6,5,4,1,7,8,2,3,9,10,11)]
yOrd <- data$dep_edu_level
k <- dim(xMat)[2]
J <- dim(as.array(unique(yOrd)))[1]
b0 <- array(rep(0, k), dim = c(k, 1))
B0 <- 1*diag(k)
d0 <- array(0, dim = c(J-2, 1))
D0 <- 0.25*diag(J - 2)
p <- 0.5

EducAtt <- quantregOR1(y = yOrd, x = xMat, b0, B0, d0, D0, burn = 1125,
                       mcmc = 4500, p, tune=1, accutoff = 0.5, TRUE)

[1] Summary of MCMC draws:

                Post Mean Post Std Upper Credible Lower Credible  Inef Factor
intercept         -3.2546   0.2175        -2.8350        -3.6798       2.3143
fam_income_sqrt    0.2788   0.0230         0.3254         0.2337       2.1396
mother_educ        0.1242   0.0190         0.1619         0.0878       1.9499
father_educ        0.1866   0.0154         0.2165         0.1578       2.3062
mother_work        0.0664   0.0821         0.2287        -0.0930       1.9349
female             0.3492   0.0786         0.5070         0.2022       1.9086
black              0.4413   0.0997         0.6400         0.2506       1.9270
urban             -0.0777   0.0971         0.1104        -0.2712       1.9019
south              0.0842   0.0880         0.2529        -0.0895       1.9153
age_cohort_2      -0.0345   0.1192         0.1963        -0.2660       1.7502
age_cohort_3      -0.0426   0.1223         0.2033        -0.2849       1.9053
age_cohort_4       0.4938   0.1212         0.7256         0.2570       1.7512
delta_1            0.8988   0.0276         0.9534         0.8461       4.6186
delta_2            0.5481   0.0313         0.6146         0.4890       3.6003

[1] MH acceptance rate: 26.8%
[1] Log of Marginal Likelihood: -4923.48
[1] DIC: 9781.91
\end{verbatim}

The posterior results\footnote{The results reported here are slightly different from those presented in Rahman (2016). This difference in results, aside from lesser number of MCMC draws, is due to a different approach in sampling from the GIG distribution. Rahman (2016) employed the ratio of uniforms method to sample from the GIG distribution (Dagpunar 2007), while the current paper utilizes the \(\texttt{rgig}\) function in the \CRANpkg{GIGrvg} package that overcomes the disadvantages associated with sampling using the ratio of uniforms method (see \CRANpkg{GIGrvg} documentation for further details). Also, see Devroye (2014) for an efficient sampling technique from a GIG distribution.} from the MCMC draws are summarized above. In the summary, we report the posterior mean and posterior standard deviation of the parameters (\(\beta_{p}, \delta_{p}\)), 95\% posterior credible interval, and the inefficiency factors. Additionally, the summary displays the MH acceptance rate of \(\delta_{p}\), the logarithm of marginal likelihood, and the DIC.

\begin{verbatim}
mcmc <- 4500
burn <- round(0.25*mcmc)
nsim <- mcmc + burn
mcmcDraws <- cbind(t(EducAtt$betadraws), t(EducAtt$deltadraws))
color_scheme_set('darkgray')
bayesplot_theme_set(theme_minimal())
mcmc_trace(mcmcDraws[(burn+1):nsim, ], facet_args = list(ncol = 3))
\end{verbatim}

\begin{figure}

{\centering \includegraphics[width=1\linewidth,height=0.5\textheight]{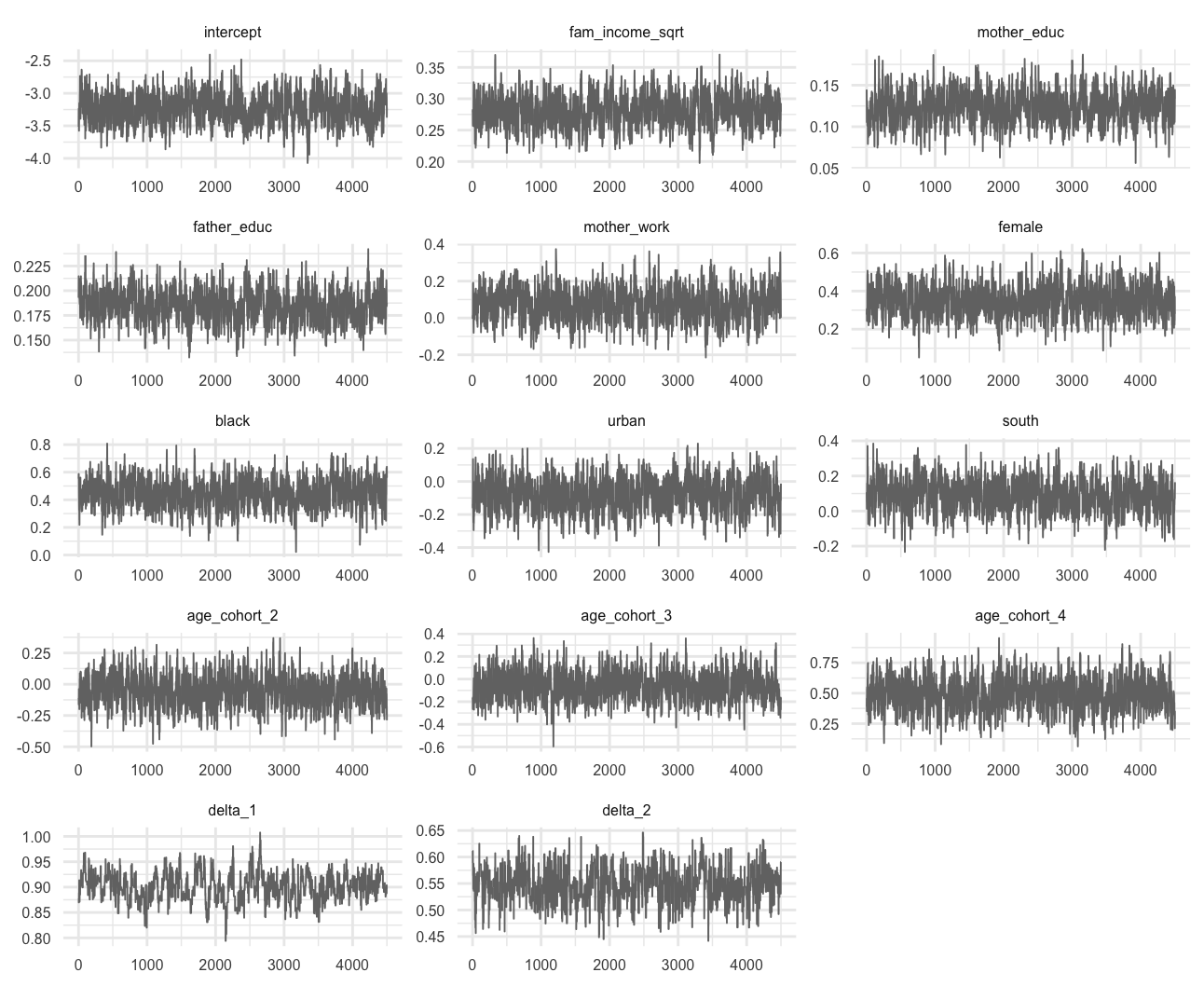} 

}

\caption{Trace plots of the MCMC draws in the educational attainment study.}\label{fig:EducTracePlot-tab-static}
\end{figure}

Figure \ref{fig:EducTracePlot-tab-static} presents the trace plots of the MCMC draws, which can be generated by loading the \CRANpkg{bayesplot} package and using the codes presented above. The purpose of trace plots is to show that the Markov chains have converged to the joint posterior distribution, such as shown in Figure \ref{fig:EducTracePlot-tab-static}. The idea is that the trace plots should show variation around a central value if the chain has converged, rather than drift without settling down.

Next, we utilize the \texttt{covEffectOR1} function to compute the average covariate effect for a \$10,000 increase in family income on the four categories of educational attainment. In general, the calculation of average covariate effect requires creation of either one or two new covariate matrices depending whether the covariate is continuous or indicator (binary), respectively. Since income is a continuous variable, a modified covariate matrix is created by adding \$10,000 to each observation of family income. This is \texttt{xMod2} in the code below and the increased income variable corresponds to \(x_{i,l}^{b}\) for \(i=1,\cdots,n\) in \protect\hyperlink{subsec:SimDataORI}{Section 4.1}. The second covariate matrix \texttt{xMod1} (in the code below) is simply the covariate or design matrix \texttt{xMat}; so with reference to \protect\hyperlink{subsec:SimDataORI}{Section 4.1}, \(x_{i,l}^{a} = x_{i,l}\) for \(i=1,\cdots,n\). When the covariate of interest is an indicator variable, then \texttt{xMod1} also requires modification as illustrated in the tax policy application.

We now call the \texttt{covEffectOR1} function and supply the inputs to get the results.

\begin{verbatim}
xMat1 <- xMat
xMat2 <- xMat
xMat2$fam_income_sqrt <- sqrt((xMat1$fam_income_sqrt)^2 + 10)
EducAttCE <- covEffectOR1(EducAtt, yOrd, xMat1, xMat2, p = 0.5, verbose = TRUE)

[1] Summary of Covariate Effect:

            Covariate Effect
Category_1          -0.0314
Category_2          -0.0129
Category_3           0.0193
Category_4           0.0250
\end{verbatim}

The results shows that at the 50th quantile and for a \$10,000 increase in family income, the probability of obtaining \(\textit{less than high school}\) (\(\textit{high school degree}\)) decreases by 3.14 (1.29) percent, while the probability of achieving \(\textit{some college or associate's degree}\) (\(\textit{college or graduate degree}\)) increases by 1.93 (2.50) percent.

\hypertarget{subsec:TaxPolicy}{%
\subsubsection{Tax Policy}\label{subsec:TaxPolicy}}

Here, the objective is to analyze the factors that affect public opinion on the proposal to raise federal taxes for couples (individuals) earning more than \$250,000 (\$200,000) per year in the United States (US). The proposal was designed to extend the Bush Tax cuts for the lower and middle income classes, but restore higher rates for the richer class. Such a policy is considered pro-growth, since it is aimed to promote economic growth in the US by augmenting consumption among the low-middle income families. After extensive debate, the proposed policy received a two year extension and formed a part of the ``Tax Relief, Unemployment Insurance Reauthorization, and Job Creation Act of 2010''.

\begin{figure}[b]

{\centering \includegraphics[width=0.4\linewidth,height=0.23\textheight]{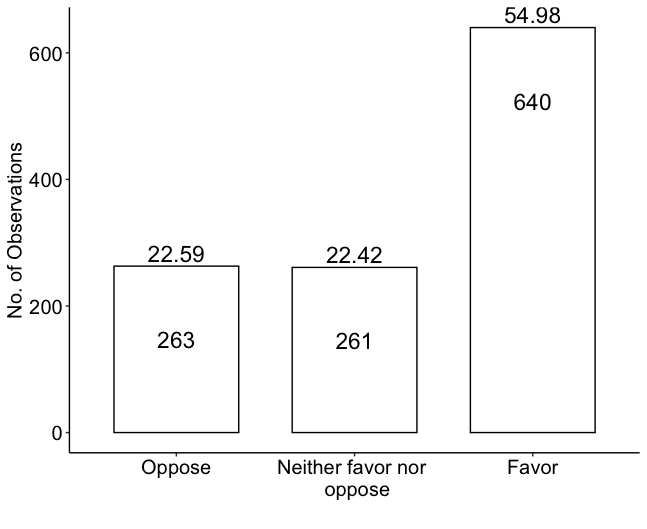} 

}

\caption{Bar chart for public opinion on tax increase. The number of responses (percentage) for each category are shown inside (at the top) of each bar.}\label{fig:TaxFreqDist-tab-static}
\end{figure}

The data for the study was taken from the 2010-2012 American National Election Studies (ANES) on the Evaluations of Government and Society Study 1 (EGSS 1) and contains 1,164 observations. The dependent variable in the model, individual's opinion on tax increase, has 3 categories: \(\textit{Oppose}\), \(\textit{Neither favor nor oppose}\), and \(\textit{Favor}\) (See Figure \ref{fig:TaxFreqDist-tab-static}). The covariates included in the model are the intercept, indicator variables for employment status, income above \$75,000, bachelors' degree, post-bachelors' degree, computer ownership, cell phone ownership, and white race.

To estimate the quantile model on public opinion about federal tax increase, we load the \CRANpkg{bqror} package, prepare the data, and provide the inputs into the \texttt{quantregOR2} function. Specifically, we define the outcome variable, covariate matrix (with covariates in order as in Rahman 2016), specify the prior distributions for \((\beta_{p}, \sigma_{p})\), choose the second cut-off value, burn-in size, number of post burn-in MCMC iterations, and values for quantile (\(p=0.5\) for this illustration) and autocorrelation cutoff.

\begin{verbatim}
library(bqror)
data("Policy_Opinion")
data <- na.omit(Policy_Opinion)
cols <- c("Intercept","EmpCat","IncomeCat","Bachelors","Post.Bachelors",
          "Computers","CellPhone", "White")
x <- data[cols]
xMat <- x[,c(1,2,3,4,5,6,7,8)]
yOrd <- data$y
k <- dim(x)[2]
b0 <- array(rep(0, k), dim = c(k, 1))
B0 = 1*diag(k)
n0 <- 5
d0 <- 8
FedTax <- quantregOR2(y = yOrd, x = xMat, b0, B0, n0, d0, gammacp2 = 3,
                      burn = 1125, mcmc = 4500, p = 0.5, accutoff = 0.5, TRUE)

[1] Summary of MCMC draws :

               Post Mean Post Std Upper Credible Lower Credible  Inef Factor
Intercept         2.0142   0.4553         2.9071         1.0959       1.4473
EmpCat            0.2496   0.2953         0.8294        -0.3270       1.6760
IncomeCat        -0.5083   0.3323         0.1329        -1.1580       1.6700
Bachelors         0.0809   0.3744         0.8569        -0.6324       1.6726
Post.Bachelors    0.5082   0.4406         1.3964        -0.3435       1.5053
Computers         0.7167   0.3509         1.4078         0.0219       1.4975
CellPhone         0.8464   0.4027         1.6191         0.0444       1.4524
White             0.0627   0.3659         0.7579        -0.6502       1.4931
sigma             2.2205   0.1442         2.5300         1.9552       2.3161

[1] Log of Marginal Likelihood: -1174.11
[1] DIC: 2334.58
\end{verbatim}

\begin{figure}

{\centering \includegraphics[width=1\linewidth,height=0.4\textheight]{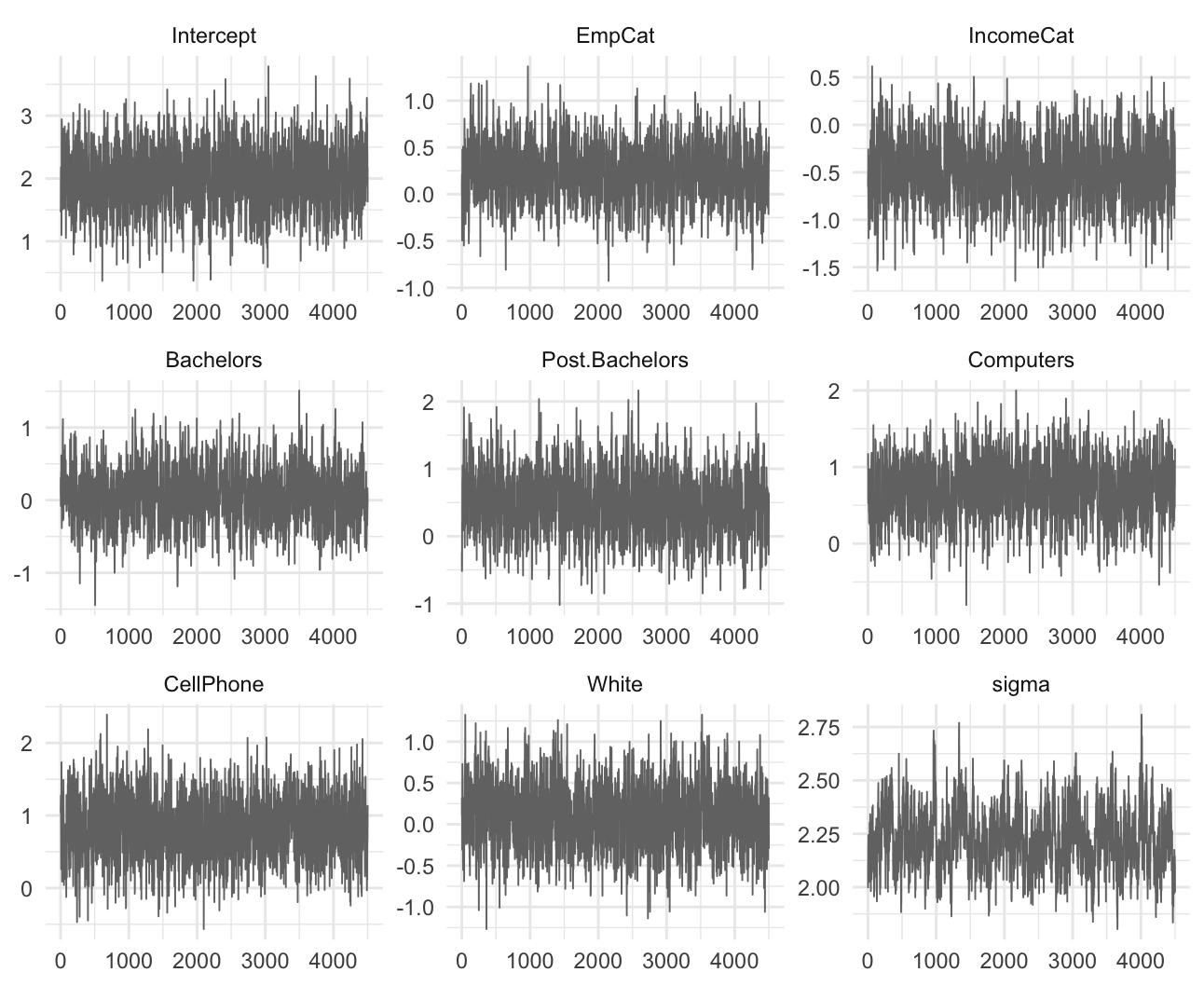} 

}

\caption{Trace plots of the MCMC draws in the tax policy study.}\label{fig:TaxTracePlot-tab-static}
\end{figure}

The results (See Footnote 1) from the MCMC draws are summarized above, where we report the posterior mean, posterior standard deviation, 95\% posterior credible interval, and the inefficiency factor of the parameters (\(\beta_{p}, \sigma_{p}\)). Additionally, the summary displays the logarithm of marginal likelihood and the DIC. Figure \ref{fig:TaxTracePlot-tab-static} presents the trace plots of the Gibbs draws, which can be generated by loading the \CRANpkg{bayesplot} package and using the codes below.

\begin{verbatim}
mcmc <- 500
burn <- round(0.25*mcmc)
nsim <- mcmc + burn
mcmcDraws <- cbind(t(FedTax$betadraws), t(FedTax$sigmadraws))
color_scheme_set('darkgray')
bayesplot_theme_set(theme_minimal())
mcmc_trace(mcmcDraws[(burn+1):nsim, ], facet_args = list(ncol = 3))
\end{verbatim}

Finally, we utilize the \texttt{covEffectOR2} function to demonstrate the calculation of average covariate effect within the \(OR_{II}\) framework. Below, we compute the average covariate effect for computer ownership (assume this is the \(l\)-th variable) on the 3 categories of public opinion about the tax policy. Here, the covariate is an indicator variable since you may either own a computer (coded as 1) or not (coded as 0). So, we need to create two modified covariate matrices. In the code snippet below, the first matrix \texttt{xMat1} and the second matrix \texttt{xMat2} are created by replacing the column on computer ownership with a column of zeros and ones, respectively. With reference to notations in \protect\hyperlink{subsec:SimDataORII}{Section 4.1} and \protect\hyperlink{subsec:SimDataORII}{Section 4.2}, \(x_{i,l}^{a} = 0\) and \(x_{i,l}^{b} = 1\) for \(i=1,\cdots,n\). We then call the \texttt{covEffectOR2} function and supply the inputs to get the results.

\begin{verbatim}
xMat1 <- xMat
xMat1$Computers <- 0
xMat2 <- xMat
xMat2$Computers <- 1
FedTaxCE <- covEffectOR2(FedTax, yOrd, xMat1, xMat2, gammacp2 = 3, p = 0.5,
                         verbose = TRUE)

[1] Summary of Covariate Effect:

            Covariate Effect
Category_1          -0.0396
Category_2          -0.0331
Category_3           0.0726
\end{verbatim}

The result on covariate effect shows that at the 50th quantile, ownership of computer decreases probability for the first category \(\textit{Oppose}\) (\(\textit{Neither favor nor oppose}\)) by 3.96 (3.31) percent, and increases the probability for the third category \(\textit{Favor}\) by 7.26 percent.

\hypertarget{conclusion}{%
\section{Conclusion}\label{conclusion}}

A wide class of applications in economics, finance, marketing, and the social sciences have dependent variables which are ordinal in nature (i.e., they are discrete and ordered) and are characterized by an underlying continuous variable. Modeling and analysis of such variables have been typically confined to ordinal probit or ordinal logit models, which offers information on the average probability of outcome variable given the covariates. However, a recently proposed method by Rahman (2016) allows Bayesian quantile modeling of ordinal data and thus presents the tool for a more comprehensive analysis and inference. The prevalence of ordinal responses in applications is well known and hence a software package that allows Bayesian quantile analysis with ordinal data will be of immense interest to applied researchers from different fields, including economics and statistics.

The current paper presents an implementation of the \CRANpkg{bqror} package -- the only package available for estimation and inference of Bayesian quantile regression in ordinal models. The package offers two MCMC algorithms for estimating ordinal quantile models. An ordinal quantile model with 3 or more outcomes is estimated by a combination of Gibbs sampling and MH algorithm, while estimation of an ordinal quantile model with exactly 3 outcomes utilizes a simpler and computationally faster algorithm that relies solely on Gibbs sampling. For both forms of ordinal quantile models, the \CRANpkg{bqror} package also provides functions to calculate the covariate effects (for continuous as well as binary regressors) and measures for model comparison -- marginal likelihood and the DIC. The paper explains how to compute the marginal likelihood from the MCMC outputs and recommends its use over the DIC for model comparison. Additionally, this paper demonstrates the usage of functions for estimation and analysis of Bayesian quantile regression with ordinal data on simulation studies and two applications related to educational attainment and tax policy. In the future, the current package will be extended to include ordinal quantile regression with longitudinal data and variable selection in ordinal quantile regression with cross section and/or longitudinal data.

\hypertarget{references}{%
\section*{References}\label{references}}
\addcontentsline{toc}{section}{References}

\hypertarget{refs}{}
\begin{CSLReferences}{1}{0}
\leavevmode\vadjust pre{\hypertarget{ref-Albert-Chib-1993}{}}%
Albert, James, and Siddhartha Chib. 1993. {``Bayesian Analysis of Binary and Polychotomous Response Data.''} \emph{{Journal of the American Statistical Association}} 88 (422): 669--79. \url{https://doi.org/10.1080/01621459.1993.10476321}.

\leavevmode\vadjust pre{\hypertarget{ref-Albert-Chib-2001}{}}%
---------. 2001. {``Sequential Ordinal Modeling with Applications to Survival Data.''} \emph{Biometrics} 57: 829--36. \url{https://www.jstor.org/stable/3068422}.

\leavevmode\vadjust pre{\hypertarget{ref-Alhamzawi-2016}{}}%
Alhamzawi, Rahim. 2016. {``Bayesian Model Selection in Ordinal Quantile Regression.''} \emph{{Computational Statistics and Data Analysis}} 103: 68--78. \url{https://doi.org/10.1016/j.csda.2016.04.014}.

\leavevmode\vadjust pre{\hypertarget{ref-Alhamzawi-Ali-Longitudinal2018}{}}%
Alhamzawi, Rahim, and Haithem Taha Mohammad Ali. 2018. {``Bayesian Quantile Regression for Ordinal Longitudinal Data.''} \emph{Journal of Applied Statistics} 45 (5): 815--28. \url{https://doi.org/10.1080/02664763.2017.1315059}.

\leavevmode\vadjust pre{\hypertarget{ref-Benoit-Poel-2010}{}}%
Benoit, Dries F., and Dirk Van den Poel. 2010. {``Binary Quantile Regression: A {B}ayesian Approach Based on the Asymmetric {L}aplace Distribution.''} \emph{Journal of Applied Econometrics} 27 (7): 1174--88. \url{https://doi.org/10.1002/jae.1216}.

\leavevmode\vadjust pre{\hypertarget{ref-Benoit-Poel-Rpackage-2017}{}}%
---------. 2017. {``Bayes{QR}: A {B}ayesian Approach to Quantile Regression.''} \emph{Journal of Statistical Software} 76 (7). \url{https://cran.r-project.org/web/packages/bayesQR/index.html}.

\leavevmode\vadjust pre{\hypertarget{ref-Bresson-etal-2021}{}}%
Bresson, Goerges, Guy Lacroix, and Mohammad Arshad Rahman. 2021. {``Bayesian Panel Quantile Regression for Binary Outcomes with Correlated Random Effects: An Application on Crime Recidivism in {Canada}.''} \emph{Empirical Economics} 60 (1): 227--59. \url{https://doi.org/10.1007/s00181-020-01893-5}.

\leavevmode\vadjust pre{\hypertarget{ref-Casella-George-1992}{}}%
Casella, George, and Edward I George. 1992. {``Explaining the {Gibbs} Sampler.''} \emph{The American Statistician} 46 (3): 167--74. \url{https://doi.org/10.1080/00031305.1992.10475878}.

\leavevmode\vadjust pre{\hypertarget{ref-Chib-Greenberg-1995}{}}%
Chib, S., and E. Greenberg. 1995. {``Understanding the {Metropolis-Hastings} Algorithm.''} \emph{The American Statistician} 49: 327--35. \url{https://doi.org/10.1080/00031305.1995.10476177}.

\leavevmode\vadjust pre{\hypertarget{ref-Chib-1995}{}}%
Chib, Siddhartha. 1995. {``Marginal Likelihood from the {Gibbs} Output.''} \emph{{Journal of the American Statistical Association}} 90 (432): 1313--21. \url{https://doi.org/10.1080/01621459.1995.10476635}.

\leavevmode\vadjust pre{\hypertarget{ref-ChibHandbook-2012}{}}%
---------. 2012. {``Introduction to Simulation and {MCMC} Methods.''} In \emph{The {O}xford Handbook of {B}ayesian Econometrics}, edited by John Geweke, Gary Koop, and Herman Van Dijk, 183--218. Oxford University Press, Oxford. \url{https://doi.org/10.1093/oxfordhb/9780199559084.013.0006}.

\leavevmode\vadjust pre{\hypertarget{ref-Chib-Jeliazkov-2001}{}}%
Chib, Siddhartha, and Ivan Jeliazkov. 2001. {``Marginal Likelihood from the {Metropolis-Hastings} Output.''} \emph{{Journal of the American Statistical Association}} 96 (453): 270--81. \url{https://doi.org/10.1198/016214501750332848}.

\leavevmode\vadjust pre{\hypertarget{ref-Chib-Jeliazkov-2006}{}}%
---------. 2006. {``Inference in Semiparametric Dynamic Models for Binary Longitudinal Data.''} \emph{{Journal of the American Statistical Association}} 101 (474): 685--700. \url{https://doi.org/10.1198/016214505000000871}.

\leavevmode\vadjust pre{\hypertarget{ref-Dagpunar-2007}{}}%
Dagpunar, John. 2007. \emph{Simulations and {M}onte {C}arlo: With Applications in Finance and {MCMC}}. John Wiley \& Sons Ltd., UK. \url{https://onlinelibrary.wiley.com/doi/book/10.1002/9780470061336}.

\leavevmode\vadjust pre{\hypertarget{ref-Davino-etal-2014}{}}%
Davino, Cristino, Marileno Furno, and Domenico Vistocco. 2014. \emph{Quantile Regression: Theory and Applications}. John Wiley \& Sons, Chichester. \url{https://doi.org/10.1002/9781118752685}.

\leavevmode\vadjust pre{\hypertarget{ref-Devroye-2014}{}}%
Devroye, Luc. 2014. {``Random Variate Generation for the Generalized Inverse {Gaussian} Distribution.''} \emph{Statistics and Computing} 24 (2): 239--46. \url{https://doi.org/10.1007/s11222-012-9367-z}.

\leavevmode\vadjust pre{\hypertarget{ref-Furno-Vistocco-2018}{}}%
Furno, Marilena, and Domenico Vistocco. 2018. \emph{Quantile Regression: Estimation and Simulation}. John Wiley \& Sons, New Jersey. \url{https://doi.org/10.1002/9781118863718}.

\leavevmode\vadjust pre{\hypertarget{ref-Gelman-etal-2013}{}}%
Gelman, Andrew, John B. Carlin, Hal S. Stern, David B. Dunson, Aki Vehtari, and Donald B. Rubin. 2013. \emph{Bayesian Data Analysis}. Chapman \& Hall, New York. \url{https://doi.org/10.1201/b16018}.

\leavevmode\vadjust pre{\hypertarget{ref-Ghasemzadeh-etal-2018-METRON}{}}%
Ghasemzadeh, Siamak, Mojtaba Ganjali, and Taban Baghfalaki. 2018. {``Bayesian Quantile Regression for Analyzing Ordinal Longitudinal Responses in the Presence of Non-Ignorable Missingness.''} \emph{METRON} 76 (3): 321--48. \url{https://doi.org/10.1007/s40300-018-0136-4}.

\leavevmode\vadjust pre{\hypertarget{ref-Ghasemzadeh-etal-2020-Comm}{}}%
---------. 2020. {``Bayesian Quantile Regression for Joint Modeling of Longitudinal Mixed Ordinal Continuous Data.''} \emph{Communications in Statistics \(-\) Simulation and Computation} 49 (2): 1375--95. \url{https://doi.org/10.1080/03610918.2018.1484482}.

\leavevmode\vadjust pre{\hypertarget{ref-Greenberg-2012}{}}%
Greenberg, Edward. 2012. \emph{Introduction to {Bayesian} Econometrics}. 2nd Edition, Cambridge University Press, New York. \url{https://doi.org/10.1017/CBO9780511808920}.

\leavevmode\vadjust pre{\hypertarget{ref-Greene-Hensher-2010}{}}%
Greene, William H., and David A. Hensher. 2010. \emph{Modeling Ordered Choices: A Primer}. Cambridge University Press, Cambridge. \url{https://doi.org/10.1017/CBO9780511845062}.

\leavevmode\vadjust pre{\hypertarget{ref-Jeliazkov-etal-2008}{}}%
Jeliazkov, Ivan, Jennifer Graves, and Mark Kutzbach. 2008. {``Fitting and Comparison of Models for Multivariate Ordinal Outcomes.''} \emph{Advances in Econometrics: {Bayesian} Econometrics} 23: 115--56. \url{https://doi.org/10.1016/S0731-9053(08)23004-5}.

\leavevmode\vadjust pre{\hypertarget{ref-Jeliazkov-Rahman-2012}{}}%
Jeliazkov, Ivan, and Mohammad Arshad Rahman. 2012. {``Binary and Ordinal Data Analysis in Economics: Modeling and Estimation.''} In \emph{Mathematical Modeling with Multidisciplinary Applications}, edited by Xin She Yang, 123--50. John Wiley \& Sons Inc., New Jersey. \url{https://doi.org/10.1002/9781118462706.ch6}.

\leavevmode\vadjust pre{\hypertarget{ref-Johnson-Albert-2000}{}}%
Johnson, Valen E., and James H. Albert. 2000. \emph{Ordinal Data Modeling}. Springer, New York. \url{https://doi.org/10.1007/b98832}.

\leavevmode\vadjust pre{\hypertarget{ref-KoenkerBook-2005}{}}%
Koenker, Roger. 2005. \emph{Quantile Regression}. Cambridge University Press, Cambridge. \url{https://doi.org/10.1257/jep.15.4.143}.

\leavevmode\vadjust pre{\hypertarget{ref-Koenker-Basset-1978}{}}%
Koenker, Roger, and G. W. Bassett. 1978. {``Regression Quantiles.''} \emph{Econometrica} 46 (1): 33--50. \url{https://doi.org/10.2307/1913643}.

\leavevmode\vadjust pre{\hypertarget{ref-Kordas-2006}{}}%
Kordas, Gregory. 2006. {``Smoothed Binary Regression Quantiles.''} \emph{Journal of Applied Econometrics} 21 (3): 387--407. \url{https://doi.org/10.1002/jae.843}.

\leavevmode\vadjust pre{\hypertarget{ref-Kozumi-Kobayashi-2011}{}}%
Kozumi, Hideo, and Genya Kobayashi. 2011. {``Gibbs Sampling Methods for {Bayesian} Quantile Regression.''} \emph{Journal of Statistical Computation and Simulation} 81 (11): 1565--78. \url{https://doi.org/10.1080/00949655.2010.496117}.

\leavevmode\vadjust pre{\hypertarget{ref-Mukherjee-Rahman-2016}{}}%
Mukherjee, Deep, and Mohammad Arshad Rahman. 2016. {``To Drill or Not to Drill? An Econometric Analysis of {US} Public Opinion.''} \emph{Energy Policy} 91: 341--51. \url{https://doi.org/10.1016/j.enpol.2015.11.023}.

\leavevmode\vadjust pre{\hypertarget{ref-PoirierBook-1995}{}}%
Poirier, Dale J. 1995. \emph{Intermediate Statistics and Econometrics}. The MIT Press, Cambridge. \url{https://mitpress.mit.edu/9780262660945/intermediate-statistics-and-econometrics/}.

\leavevmode\vadjust pre{\hypertarget{ref-Rahman-2016}{}}%
Rahman, Mohammad Arshad. 2016. {``Bayesian Quantile Regression for Ordinal Models.''} \emph{Bayesian Analysis} 11 (1): 1--24. \url{https://doi.org/10.1214/15-BA939}.

\leavevmode\vadjust pre{\hypertarget{ref-Rahman-Karnawat-2019}{}}%
Rahman, Mohammad Arshad, and Shubham Karnawat. 2019. {``Flexible {Bayesian} Quantile Regression in Ordinal Models.''} \emph{Advances in Econometrics} 40B: 211--51. \url{https://doi.org/10.1108/S0731-90532019000040B011}.

\leavevmode\vadjust pre{\hypertarget{ref-Rahman-Vossmeyer-2019}{}}%
Rahman, Mohammad Arshad, and Angela Vossmeyer. 2019. {``Estimation and Applications of Quantile Regression for Binary Longitudinal Data.''} \emph{Advances in Econometrics} 40B: 157--91. \url{https://doi.org/10.1108/S0731-90532019000040B009}.

\leavevmode\vadjust pre{\hypertarget{ref-Spiegelhalter-etal-2002}{}}%
Spiegelhalter, David J., Nicola G. Best, Bradley P. Carlin, and Angelika Van Der Linde. 2002. {``Bayesian Measures of Model Complexity and Fit.''} \emph{{Journal of the Royal Statistical Society -- Series B}} 64 (4): 583--639. \url{https://doi.org/10.1111/1467-9868.00353}.

\leavevmode\vadjust pre{\hypertarget{ref-Tian-etal-2021}{}}%
Tian, Yu-Zhu, Man-Lai Tang, Wai-Sum Chan, and Mao-Zai Tian. 2021. {``Bayesian Bridge-Randomized Penalized Quantile Regression for for Ordinal Longitudinal Data, with Application to Firm's Bond Ratings.''} \emph{Computational Statistics} 36: 1289--319. \url{https://doi.org/10.1007/s00180-020-01037-4}.

\leavevmode\vadjust pre{\hypertarget{ref-Yu-Moyeed-2001}{}}%
Yu, Keming, and Rana A. Moyeed. 2001. {``Bayesian Quantile Regression.''} \emph{Statistics and Probability Letters} 54 (4): 437--47. \url{https://doi.org/10.1016/S0167-7152(01)00124-9}.

\end{CSLReferences}

\bibliography{rahman.bib}

\address{%
Prajual Maheshwari\\
Quantitative Researcher, Ogha Research\\%
2123, 14th Main Road, HAL 3rd Stage, Kodihalli, Bengaluru, Karnataka, India\\
\url{https://prajual.netlify.app}\\%
\href{mailto:prajual1391@gmail.com}{\nolinkurl{prajual1391@gmail.com}}%
}

\address{%
Mohammad Arshad Rahman\\
Department of Economic Sciences, Indian Institute of Technology Kanpur\\%
Room 672, Faculty Building, Indian Institute of Technology Kanpur, India\\
\url{https://www.arshadrahman.com}\\%
\textit{ORCiD: \href{https://orcid.org/0000-0001-8434-0042}{0000-0001-8434-0042}}\\%
\href{mailto:marshad@iitk.ac.in}{\nolinkurl{marshad@iitk.ac.in}}, \href{mailto:arshadrahman25@gmail.com}{\nolinkurl{arshadrahman25@gmail.com}}%
}

\end{article}

\end{document}